\documentclass[journal]{IEEEtran}

\usepackage{cite}
\usepackage{amsmath,amssymb,amsfonts}
\usepackage{algorithmic}
\usepackage{graphicx}
\usepackage{textcomp}
\usepackage[table]{xcolor}
\def\BibTeX{{\rm B\kern-.05em{\sc i\kern-.025em b}\kern-.08em
    T\kern-.1667em\lower.7ex\hbox{E}\kern-.125emX}}
\usepackage{orcidlink}    
\usepackage{booktabs,multirow,bigstrut}
\usepackage[nameinlink,noabbrev,capitalise]{cleveref}
\usepackage{orcidlink}

\begin{document}
\title{Towards Neuromorphic Compression based Neural Sensing for Next-Generation Wireless Implantable Brain Machine Interface}
%%%%%%%%%%%%%%%%%%%%%%%%%%% Authors list %%%%%%%%%%%%%%%%%%%%%%%
\author{
Vivek~Mohan~\orcidlink{0000-0002-0248-6417},~\IEEEmembership{Member,~IEEE},
Wee~Peng~Tay~\orcidlink{0000-0002-1543-195X},~\IEEEmembership{Senior Member,~IEEE},  and 
Arindam~Basu~\orcidlink{0000-0003-1035-8770},~\IEEEmembership{Senior Member,~IEEE}

\thanks{V. Mohan and W. P. Tay are with the School of Electrical and Electronic Engineering, Nanyang Technological University, Singapore. (e-mail: vivekmoh001@e.ntu.edu.sg, wptay@ntu.edu.sg)}

\thanks{A. Basu is with the Department of Electrical Engineering, City University of Hong Kong. (e-mail: arinbasu@cityu.edu.hk)}

\thanks{Note: This work has been submitted to the IEEE for possible publication. Copyright may be transferred without notice, after which this version may no longer be accessible.}
}

%%%%%%%%%%%%%%%%%%%%%%%%%%%% The paper headers %%%%%%%%%%%%%%%%%%%%
\markboth{IEEE Transactions Paper,~Vol.~XX, No.~XX, Month~XXXX}%
{Mohan \MakeLowercase{\textit{et al.}}:Neuromorphic Compression based Neural Sensing for Next-Gen Wireless implantable-BMI}

\maketitle

%%%%%%%%%%%%%%%%%%%%%%%%%%%% INCLUDE abstract.tex and Keywords %%%%%%%%%%%%%%%%
\begin{abstract}
  This work introduces a neuromorphic compression based neural sensing architecture with address-event representation inspired readout protocol for massively parallel, next-gen wireless iBMI. The architectural trade-offs and implications of the proposed method are quantitatively analyzed in terms of compression ratio and spike information preservation. For the latter, we use metrics such as root-mean-square error and correlation coefficient between the original and recovered signal to assess the effect of neuromorphic compression on spike shape. Furthermore, we use accuracy, sensitivity, and false detection rate to understand the effect of compression on downstream iBMI tasks, specifically, spike detection. We demonstrate that a data compression ratio of $50-100$ can be achieved, $5-18\times$ more than prior work, by selective transmission of event pulses corresponding to neural spikes. A correlation coefficient of $\approx0.9$ and spike detection accuracy of over $90\%$ for the worst-case analysis involving $10K$-channel simulated recording and typical analysis using $100$ or $384$-channel real neural recordings. We also analyze the collision handling capability and scalability of the proposed pipeline.
\end{abstract}


\begin{flushleft}\label{AbbreviationsList}
\footnotesize{
\textbf{List of Abbreviations- } 

\begin{table}[h]
% \begin{tabular}{>{\color{red}}l >{\color{red}}l}
\begin{tabular}{ll}
% \begin{supertabular}{ll}
iBMI    & Implantable Brain Machine Interface\\
ADC     & Analog to Digital Converter\\
ADM      & Asynchronous Delta Modulator\\
SPDWOR    & Wired-OR Readout\\
DVS    & Dynamic Vision Sensor\\
AER    & Address event representation\\
APM    & All Pulse Mode\\
PCM    & Pulse Count Mode\\
NHP    & Non-human Primate\\
RMSE    & Root Mean Square Error\\
CC    & Correlation Coefficient\\
$\mathrm{Th_{ON(OFF)}}$     & ON/OFF Pulse Generation Threshold\\
$\mathrm{Th_{spd}}$     & Spike Detection Threshold\\
SPD     & Spike Detection\\
AT-SPD     & Absolute-threshold based SPD\\
NEO-SPD     & Non-linear Energy Operator based SPD\\
TDR     & Transmission Data Rate\\
CR     & Compression Ratio
\end{tabular}
\end{table}

% \begin{table}[H]
% \begin{tabular}{ll}
% \\
% RP      & Region Proposal                                                    \\
% HIST RP & Histogram RP                                                       \\
% CCL RP  & Connected Component Labeling RP                                    \\
% 1B1C    & 1-bit 1-channel image                                              \\
% 1B2C    & 1-bit 2-channels image                                             \\
% BB      & Bounding Box                                                       \\
% NNDC RP & Neural Network Detector plus Classifier RP                         \\
% GT      & Ground Truth                                                       \\
% IoU     & Intersection-over-Union                                            \\
% AUC     & Area Under Curve                                                   \\
% NMS     & Non-Maximal Suppression                                            \\
% EvFT    & Event Feature Tracker based RP \\
% EBBINN  & EBBI + NNDC RP                                                    
% % \end{supertabular}
% \end{tabular}
% \end{table}
}
\end{flushleft}

\IEEEpeerreviewmaketitle
%%%%%%%%%%%%%%%%%%%%%%%%%%%% Include section-wise files %%%%%%%%%%%%%%
\section{Introduction}
\begin{figure*}[t!]
\centering
\resizebox{0.95\textwidth}{!}{
\includegraphics[width=\textwidth]{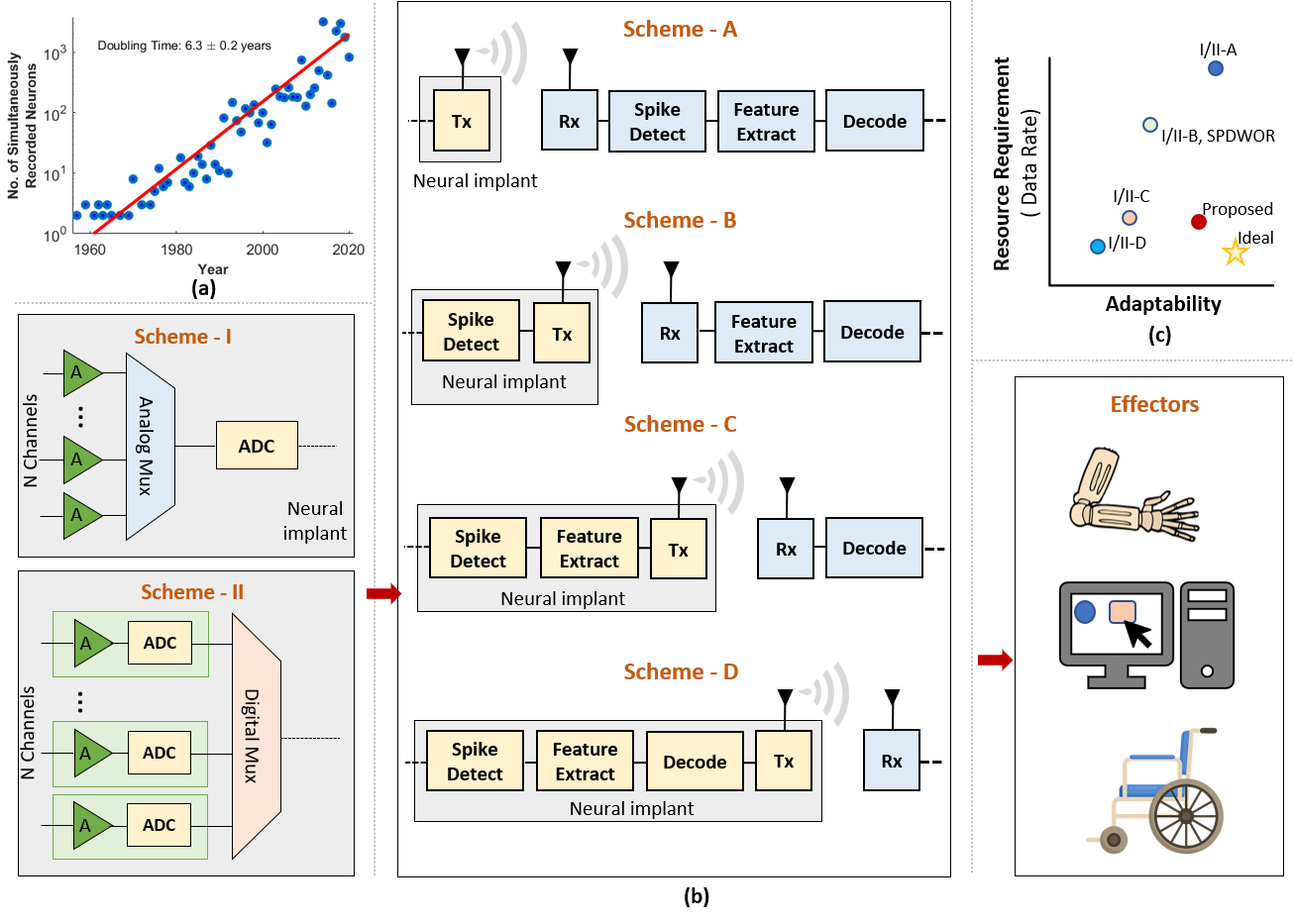}
}
\caption{(a) Studies in \cite{electrodeScaling} observed a Moore's law like doubling of the number of simultaneously recorded neurons. This trend indicates the need to develop neural processing systems that scale well with electrode count while assuring robust performance within the allowable power, memory, and data rate budget. (adapted from \cite{stevenson_2013}) (b) Block diagram of typical iBMI systems involving different schemes for multiplexing the analog/digital signals followed by different options to transmit the recorded signal, ranging from transmission of all recorded signals to transmitting decoded signals.  (c) Trade-off plot, putting into perspective the existing iBMI systems and the proposed pipeline.}
\label{fig:Figure1}
\end{figure*}

%The next generation of implantable brain machine interfaces (Nx-iBMI) intended for assistive technologies such as prosthetic arms are expected to support parallel recording from thousands of electrodes in order to improve iBMI performance and enable sophisticated control of effectors. It is also necessary to consider implementing Nx-iBMI as wireless transcutaneous implants to reduce the risk of infection, enhance aesthetics and user mobility. 
\IEEEPARstart{A}{dvances} in neurotechnology in recent years has allowed partial restoration of lost sensory capabilities such as vision\cite{VisionBMI}, speech\cite{speechBMI2023} and touch\cite{touchBMI} through stimulation, and limited motor capabilities in people suffering from motor impairment or paralysis. At the same time, a variety of neural sensors such as electroencephalography (EEG), electrocorticography (ECoG) and intracortical electrode based implantable brain-machine interface (iBMI) have demonstrated promising results for clinical applications\cite{clinicalBMI}. A typical implementation of an iBMI system involves recording neural activity through a microelectrode array followed by amplification, filtering, and spike-detection stages to capture the action potentials which then may be decoded on- or off-chip to operate and control effectors such as prosthetic arm, computer cursor, mobility devices, etc., as shown in Fig. \ref{fig:Figure1}.

Recent works using iBMI have enabled brain to handwritten text \cite{Brain2Text}, brain-based speech synthesis \cite{speechBMI}, and therapies for epilepsy via deep brain stimulation and mental disorders \cite{MentalBMI}. Despite the compelling advances in the direction of futuristic BMI assistive technologies, the efficacy of such systems is limited by the number of recording channels. It, therefore, becomes necessary for dexterous next generation of implantable brain-machine interfaces (Nx-iBMI) to support parallel recording from thousands of electrodes in order to improve iBMI performance and enable more sophisticated control of assistive technologies such as prosthetic arms. It is also necessary to consider implementing Nx-iBMI as wireless transcutaneous implants to reduce the risk of infection, enhance aesthetics, and user mobility, and allow stable recording for longer durations. Increasing the cellular coverage, i.e., the number of simultaneously recorded neurons, allows the study of granular neuronal interactions and cooperation like never before and potentially treating neural disorders or even enhancing sensory perception. With advances in semiconductor technologies, the general direction toward increasing neural signal resolution has resulted in the development of neural interfaces with higher channel counts \cite{19584Electrodes,5120Electtrodes,59760Electrodes, Argo} and thereby creating a trend of Moore's law like doubling of the number of simultaneously recorded neurons about every $6.3$ years as shown in Fig. \ref{fig:Figure1}(a). 

One of the main issues that come up with increasing the channel count is data bandwidth limitations and power dissipation, especially with wireless transmission. To illustrate the challenge with electrode scaling in neural implants, consider an example of Nx-iBMI that would have about $10,000$ channels following Moore's law like scaling of neural electrodes as shown in Fig.~\ref{fig:Figure1}(a). A major hurdle for the interface would be digitizing the massive amount of data (10 bits/sample $\times 30$Ks/sec $\times 10,000$ channels = $3$~Gbps of neural data) and transmitting it off-chip (as done in Schemes I-A and II-A in Fig. \ref{fig:Figure1}(b)). The power budget is constrained by the maximum allowable thermal power dissipation of  $\approx80$~mW/cm$^2$, and the temperature increase in the neural interface is restricted below $ 0.5^\circ$C  to prevent damage to brain tissue\cite{bmiTemperature}. 
%Considering the case of $10,000$ channels with a maximum allowable heat dissipation of $10$ mW to prevent tissue damage, the power budget per channel is about $1$ $\mu$W. 
This points to the need for data compression in the implant to satisfy the increased channel count requirements of the future and to keep the transmission power low. Another issue relates to the wiring required to access numerous electrodes in a limited area, necessitating some form of multiplexing. In this work, we analyze a neuromorphic event-driven neural front-end that can potentially address both these issues.

% \begin{comment}
% Existing iBMI systems commonly used in neuroscience studies, typically record and transmit high-resolution wide-bandwidth signals through wired cables to ensure high reliability for offline processing, resulting in extensive power and bandwidth overheads. This approach of transmitting all signals is difficult to scale in wireless iBMIs beyond data rates of a few tens of Mbps due to increased bit-error rates and low run-time between battery charge cycles. One way to address the problem is by transmitting data that carry salient information related to the spikes while discarding the unwanted data, thereby trading off a very high signal-to-noise ratio in the front-end stages for robust, resource-efficient iBMI systems. Considering the sparsity of neural spikes, it becomes necessary to implement novel low-power neural data compression and analyze their efficacy for Nx-BMI.
% \end{comment}
The rest of the paper is organized as follows. The following section discusses some of the related works and lists the key contributions of this work. Section \ref{sec:Methodology} describes the proposed neuromorphic compression pipeline. Section \ref{sec:Results} presents the trade-off between compression and performance for the neuromorphic compression pipeline and compares it with relevant conventional and novel methods from previous works. This is followed by a section \ref{sec:Discussion} that discusses the main results and shows that our approach is scalable and yields expected results with two real datasets. Finally, we summarize our findings and conclude in the last section.
\section{Related Work and Contribution}
Various research works have made progressive improvements in the iBMI sensing architecture or processing flow. Following are some popular methods that try to address the issues that arise with increasing electrode count on neural recording arrays:
\subsection{Sub-array digitization}
As the number of recording channels increases, traditionally used analog multiplexing schemes (as shown in Fig. \ref{fig:Figure1}(b) Scheme-I) become prone to noise and interference, therefore multiplexing digital signals becomes a preferred scalable solution. Newer neural recording arrays such as Neuropixel \cite{5120Electtrodes}, Neuralink \cite{Neuralink}, Argo \cite{Argo} and $59$K in-vitro electrodes \cite{59760Electrodes} attempt to tackle the power and data-rate limitations by either using programmable switch matrix or by implementing on-chip multiplexing for sub-array digitization, which is a hybrid between the Schemes I and II in Fig. \ref{fig:Figure1}(b) involving multiplexing of a group of recording electrodes in the analog domain before digitization, thereby making it feasible for only about $3-4\%$ of the electrodes  ($2,048$ channels for Argo and $384$ channels for Neuropixel) that can be addressed simultaneously. \cite{59760Electrodes,19584Electrodes} integrate the ADC under a neuro-pixel while increasing the electrode count. This allows full-frame readout from all electrodes besides ensuring high-SNR, robust digital readout from an arbitrarily selectable subset of electrodes. They however neither address the data deluge post-digitization nor provide compression strategies to realize scalable, low-power, wireless neural implants.
\subsection{On-chip compression}
Analog techniques such as spatial compression before digitization and compression via superposition were proposed in \cite{cs_2013,cs_2018}. However, they either involve large complexity hardware-intensive algorithms for the front end or are limited by noise summation, which limits the scalability of such neural implants. Compression schemes such as compressive sensing \cite{CompressiveSensing} and autoencoder \cite{AutoEncoder} also fall short of the requirements to meet the available wireless data rates with increasing electrode count. On-chip spike sorting \cite{Liu16,Liu17,zz22,onlineSpikeSort2023} and compression \cite{Pagin17} have been proposed to reduce the transmitted data (as shown in Fig. \ref{fig:Figure1}(b) Scheme-I/II followed by Scheme-C). However, these do not tackle the issue of multiplexing and digitization in the front end of the neural implant sufficiently.
\subsection{Wired-OR readout (SPDWOR)}
An interesting image sensor inspired technique, presented in \cite{SPDWOR}, exploits the spatio-temporal sparsity of neural signals to simultaneously achieve compression and digital multiplexing with wired-OR interactions. It provides a lossy compression by discarding samples in the baseline region via wired-OR contention among pixels retaining samples potentially corresponding to spikes. When multiple pixels compete for access to the limited wires, collision occurs, and more than one row/column decoders are activated, resulting in no unique decoding solution for recovery of quantized voltages. Such collisions may also occur for spike signals depending on the choice of the baseline, especially when neural firing is correlated-- leading to loss of spike data. Reducing the loss of valid samples to collision would require a more complex wiring configuration and address decoding schemes.
\subsection{Activity dependent processing}
Data rate and power-sensitive methods propose transmitting a binary train of spike information indicating the presence or absence of spikes in time bins\cite{naturePowerSaving} for real-time clinically viable iBMI. While these methods reduce power consumption and data rate by an order of magnitude, they preclude features such as spike shapes for tasks such as spike sorting. However, the performance of iBMI systems employing such compression with increasing electrode count on the neural interface array is still unknown. Some other works such as \cite{ShoebDatou, ZZ_FRM_SPD} advocate the integration of decoders in the implant (as shown in Fig. \ref{fig:Figure1}(b) Scheme-I/II followed by Scheme-D). While this addresses the problem convincingly, it also limits the implant to specific tasks, reducing its adaptability in the future \cite{Shaikh2022}. 
\subsection{Neuromorphic compression}
Limitations imposed by power and bandwidth with electrode scaling on neural interfaces could be addressed effectively with the promising low-power neuromorphic approach as demonstrated previously for sensory applications such as event-based vision \cite{dvs_tobi}, audio \cite{SiliconEar}, tactile \cite{Tactile}, and olfactory \cite{SiliconNose} sensors. This approach essentially `morphs' the initial sensory information processing stages of biological sensors/receptors into VLSI chips using a combination of efficient analog and digital circuitry to mimic their asynchronous spatio-temporal activity. A neuromorphic event-driven neural recording approach was first proposed in \cite{NeuromorphicRecSystem} for iBMI and more recently implemented in \cite{NSS_PNS, AdaptiveADM, NeuroBMI}. The neuromorphic approach can also provide benefits of digital multiplexing due to integration with address event representation (AER) circuits and digitizing/communicating data only during spikes by virtue of in-pixel thresholding. The AER-based handling of events could potentially address the shortcomings of data loss due to collisions in SPDWOR configuration. Works in \cite{Liu16, NSS_PNS, NeuroBMI} introduce a pipeline involving spike train generation using an analog-to-spike conversion block on-chip decoding using a spiking neural network, which limits its usability to specific tasks. While \cite{NSS_PNS} uses a round-robin arbitration based AER scheme, \cite{Liu16,NeuroBMI} does not have a collision handling strategy that could scale with the size of the neural electrode. None of the aforementioned neuromorphic/event-based neural recording schemes investigate the gains from using the neuromorphic compression in terms of reduction in data rate, nor assess its impact on iBMI performance. The fidelity of the compressed signal and the effect of collisions in the AER block of a neuromorphic compressive iBMI on task-specific performance when scaling to large Nx-iBMI systems has hitherto been unexplored.

In this work, we investigate the architectural trade-offs of neuromorphic compression of neural signals for iBMI inspired by asynchronous event detection image sensors (AEDIS) such as the dynamic vision sensor (DVS) using next-generation large neuromorphic neural recording systems for simulated and realistic neural recordings. We make the following contributions:
\begin{itemize}
    \item We explore the use of address event representation (AER) based readout with a pair of handshaking signals (request and acknowledgment) for elegant collision management to prevent loss of detected pulses by delaying them.
    \item We assess the extent and effect of neuromorphic compression on spike shape and spike detection performance for two modes of transmission - `All pulse mode' (APM) and `Pulse count mode' (PCM) quantitatively with a set of standard evaluation metrics and compare them with other popular methods of transmitting neural signals.
    \item We evaluate task-specific fidelity, specifically spike shape preservation and spike detection performance, for APM and PCM pipelines and investigate the performance variation across different synthetic and real datasets (100-channel non-human primate, i.e., NHP and 384-channel Neuropixel recordings).
    \item We assess the collision handling capability and effect of collision on signal recovery.
    \item We discuss the scalability of the proposed method to a higher number of recording channels and the potential to compress the proposed read-out further.
\end{itemize}
An initial version of this work was presented in \cite{Vivek2023} where the proposed neuromorphic compression based neural sensing pipeline was analyzed using a single-channel synthetic recording. This work extends the findings for higher channel count recordings including real neural recordings from non-human primates and mice to understand the scalability of the proposed pipeline, besides investigating its collision handling capability and exploring options for further compression. The neuromorphic iBMI event dataset obtained from the simulation of the proposed pipeline is made publicly available at: \textcolor{blue}{ \emph{\url{https://sites.google.com/view/brainsyslab/neuromorphic-ibmi-dataset}}}.
\begin{figure*}[t!]
\centering
\resizebox{0.98\textwidth}{!}{
\includegraphics[width=\textwidth]{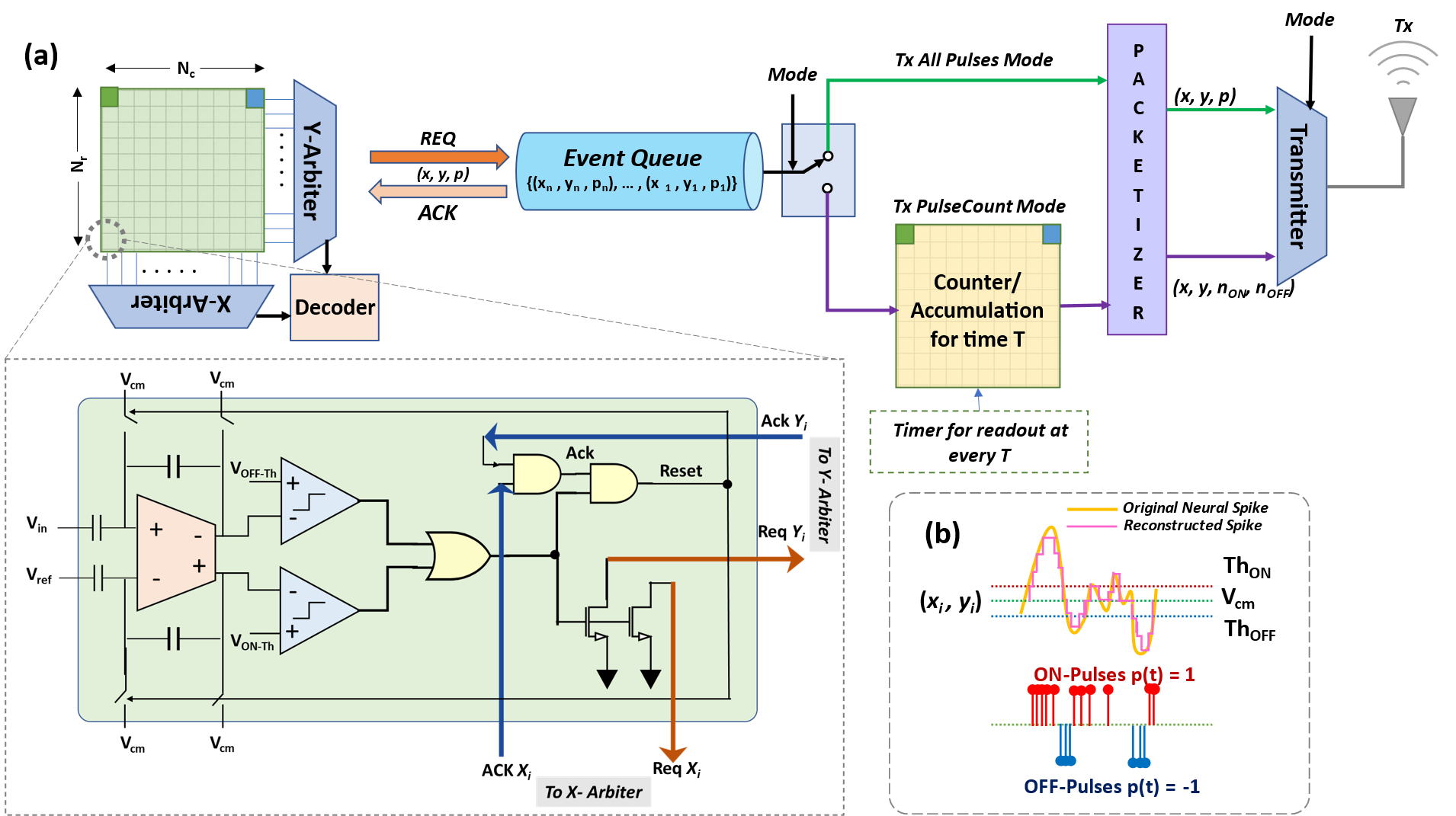}
}
\caption{(a) Neuromorphic compression based neural sensing amplifier inspired from address event representation (AER) protocol used in event-based cameras that output ON/OFF pulses only when the input exceeds the ON/OFF threshold. The circuit implementation can be similar to the implementation in \cite{NeuromorphicRecSystem}. Generated pulses can be packetized and wirelessly transmitted in `All Pulse Mode' (APM) or `Pulse Count Mode' (PCM) (b) The pulse train generated and the reconstructed waveform, which approximates the original neural spike signal.}
\label{fig:Figure2}
\end{figure*}
\section{Methodology}
\label{sec:Methodology}
The firing rate of a biological neuron is $\approx1-200$Hz. Combined with an approximate spike duration of $1-2$ ms, this implies that biological spikes occupy a small fraction of samples in neural recordings. This temporal sparsity of spikes allows the use of an AER-based readout strategy that leads to high compression rates in large iBMI interfaces.
\subsection{Overview of the Proposed Pipeline}
A neuromorphic compression based neural sensing system as shown in Fig.~\ref{fig:Figure2}(a) is proposed to consist of a DVS-pixel-like ON and OFF threshold crossing detection circuit \cite{dvs_tobi} integrated into each cell, similar to the implementation in \cite{NeuromorphicRecSystem}. Generally, the front end comprises a capacitive low-noise amplifier with gain $A_1$ followed by a second programmable gain stage with gain $A_2$. The total gain of the amplifier stages is typically $\approx 200-1000$. This is followed by an asynchronous delta modulator to generate an output $\mathrm{V_{mod}}$ and pulses $p$ (+1/ON and -1/OFF). The AER readout strategy is facilitated by arbitration logic, address decoders, and a pair of handshaking signals for event readout. The readout events are then packetized depending on the mode of operation and transmitted wirelessly for further downstream processing of the neural event stream. The following subsections elaborate on the working principle of some of the main blocks of the proposed pipeline.

\subsection{Event Generation}
Each pixel consists of an asynchronous delta modulator typically composed of an input operational transconductance amplifier (OTA) with a capacitive divider gain stage, a pair of comparators, and inverters that generate the pulse train commonly referred to as `events'. ON ($\mathrm{+1}$) and OFF ($\mathrm{-1}$) pulses/events, are produced when the change in the amplified version of the input signal ($\mathrm{V_{in}}$) increases above the reference voltage, $\mathrm{V_{ref}}$, by $\mathrm{Th_{ON}}$ or decreases below $\mathrm{V_{ref}}$ by $\mathrm{Th_{OFF}}$ respectively. The pulse generation process can be described as follows: 
%\red{[I don't understand why $\mathrm{V_{mod}(t)}=\mathrm{V_{cm}}$ appears in both cases. How about the third case? Your formula is also inconsistent with the discussion before this, which uses $\mathrm{V_{ref}}$. Somehow,$\mathrm{V_{in}} > \mathrm{V_{ref}} + \mathrm{Th_{ON}}$ is equivalent to $\mathrm{V_{mod}(t)}>\mathrm{Th_{ON}}$?]}
\begin{gather}
\label{eq:pulse}
\frac{\mathrm{dV_{mod}}}{\mathrm{dt}}=A_1A_2\frac{\mathrm{dV_{in}}}{\mathrm{dt}}\notag\\
  p\mathrm{(t)} = 
\begin{cases} 
~~1,~\mathrm{V_{mod}(t+\delta)}=\mathrm{V_{cm}},~\text{if}~\mathrm{V_{mod}(t)}>\mathrm{Th_{ON}}, \\
-1,~\mathrm{V_{mod}(t+\delta)}=\mathrm{V_{cm}},~\text{if}~\mathrm{V_{mod}(t)}<\mathrm{Th_{OFF}}, \\
~~0,~\mathrm{V_{mod}(t+\delta)}=\mathrm{V_{mod}(t)}+\frac{\mathrm{dV_{mod}}}{\mathrm{dt}}\delta,~\text{otherwise,}
\end{cases}
\end{gather}
% \begin{gather}
% \label{eq:pulse}
% \frac{\mathrm{dV_{mod}}}{\mathrm{dt}}=A_1A_2\frac{\mathrm{dV_{in}}}{\mathrm{dt}}\notag\\
% \text{if}~(\mathrm{V_{mod}(t)}>\mathrm{Th_{ON}}),p\mathrm{(t)}=1, \mathrm{V_{mod}(t)}=\mathrm{V_{cm}}\notag\\
% \text{else~if}~ (\mathrm{V_{mod}(t)}<\mathrm{Th_{OFF}}),p\mathrm{(t)}=-1,\mathrm{V_{mod}(t)}=\mathrm{V_{cm}}\notag\\
% else,~p(t)=0
% \end{gather}
where $\mathrm{V_{cm}}$ is the common mode voltage of the ADM. An example waveform with a spike reconstruction is shown in Fig.~\ref{fig:Figure2}(b). Other implementations\cite{chenyi_folding} have also combined this function in one stage, where the signal reconstruction method was different. An enhanced adaptive version of the asynchronous delta modulator, as introduced in \cite{AdaptiveADM}, could be used to modulate and minimize the event generation rate by following the amplitude and noise characteristics of the input signal. This could effectively suppress event generation due to noise or abnormalities in the baseline region. For the purpose of simplicity, the event generation block in this work is assumed to contain a basic asynchronous delta modulator, as implemented in \cite{NeuromorphicRecSystem}.
\subsection{AER-based Readout}
Upon generation of ON/OFF events, the $\mathrm{Req}$ signal is generated and, an additional logic asserts the $\mathrm{Ack}$ signal once the event is readout, which in turn resets the pixel. The reset state will be held for a `refractory period' determined by the values of the capacitor and the bias voltage in the pixel circuitry. The number of pulses generated by each pixel depends on the amplitude and frequency of the input signal, and also on the parameters of the delta modulator - ON/OFF pulse generate thresholds and refractory period. The pixels across the rows and columns of the electrode array are tied via a wired-OR connection to pass the read request (Req) to the readout block. Row $\mathrm{(Req~Y_i)}$ and column $\mathrm{(Req~X_i)}$ request lines from each of the neural recording pixels are connected to the row (Y) and column (X) arbiter similar to DVS \cite{dvs_tobi}. For realizing an AER-like readout and managing collision efficiently, we use a toggle tree fair X, Y-arbiter \cite{toggleArbiter} to decide the sequence of readout when pulse read requests are generated by multiple electrodes simultaneously. As in any AER system, time represents itself and events are generated asynchronously. The address $\mathrm{(x_{i},y_{i})}$ and polarity ($\mathrm{p_{i}}$) of the generated pulses (ON/$\mathrm{+1}$ or OFF/$\mathrm{-1}$) are communicated to the next stage. Acknowledgment signals ($\mathrm{Ack~Y_i}$ and $\mathrm{Ack~X_i}$) are then sent from the X, Y-arbiters to the cell ($\mathrm{x_i,y_i}$) after read out, to reset its amplifier for continuing the delta compression.
\subsection{Event Packaging}
The proposed neuromorphic scheme allows the neural data to be encoded as a train of asynchronous binary ON/OFF pulses, instead of producing an n-bit word for each sample at the output of the ADC. Events from the event stream are packetized depending on the mode of transmission before they are transmitted. In this work, we investigate two modes of operation: APM and PCM. In APM, all the generated pulses are transmitted off-chip asynchronously through a wireless link, with the event pulse packetized to contain the address and polarity of the event, $\mathrm{(x_{i},y_{i},p_{i})}$. In contrast, PCM involves electrode-wise accumulation of events with fixed-time bins. If the width of the bin is chosen to be `n'-sampling intervals long, then it is denoted as PCM`n' (e.g. PCM1, PCM2, and PCM4 have bin-widths equal to 1, 2, and 4 times the traditional sampling interval in neural recording systems respectively). The ON and OFF Events generated by an electrode at $\mathrm{(x_{i},y_{i})}$ location in PCM mode are packetized as the ON and OFF event counts -  $\mathrm{n_{ON}}$ and $\mathrm{n_{OFF}}$ as $\mathrm{(x_{i},y_{i},n_{ON},n_{OFF})}$. For a bin width of duration $\mathrm{t_{b}}$, in PCM the ON/OFF events are accumulated as follows:
\begin{flalign}
    \label{eq:pcm}
    \mathrm{n_{ON(OFF)}}=\sum_{\mathrm{t_{b}}} | p\mathrm{(t)}|_{p(t)=1(-1)}
\end{flalign} The number of bits per APM event is, $1 + \log_{2}\mathrm{N_r} + \log_{2}\mathrm{N_c}$ whereas the number of bits per PCM event is, $\mathrm{n_{ON}}+ \mathrm{n_{OFF}}+ \log_{2}\mathrm{N_r} + \log_{2}\mathrm{N_c}$ where $\mathrm{N_r}\times\mathrm{N_c}$ is the size of the electrode array. In both modes, nothing is transmitted when no events are generated. At the receiver end, the electrode signals can be recovered by the accumulation of pulses depending on their polarity by stair-step reconstruction or directly processed using spiking neural networks, for further processing downstream processing such as spike detection, spike sorting, decoding, etc. 
\section{Results}
\label{sec:Results}
\subsection{Simulation Setup}
\begin{figure}[t]
\centering    
\includegraphics[width=0.98\columnwidth]{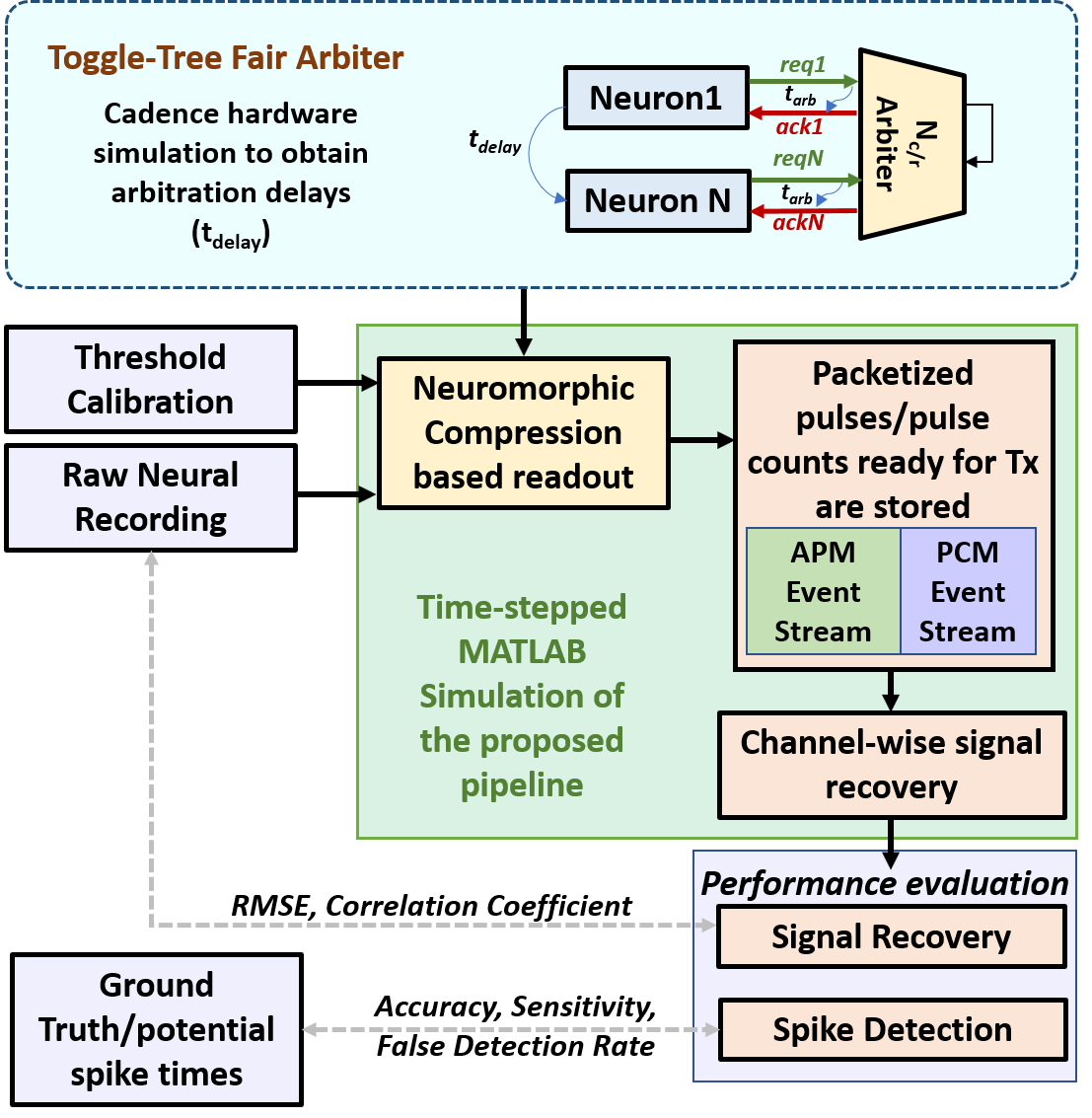}
 \caption{Block diagram of the data processing pipeline used to generate the simulation results for the proposed neuromorphic compression based neural sensing system.}
\label{fig:Figure3}
\end{figure}

Simulations were performed using the data processing pipeline shown in Fig.\ref{fig:Figure3}. In order to incorporate realistic arbitration delay when simulating for larger electrode count, we simulated a toggle-tree fair arbiter as introduced in \cite{toggleArbiter} in a $65$ nm CMOS process using Cadence Virtuoso to determine the arbitration delay. The arbitration delay ($\mathrm{t_{arb}}$) was estimated to be in the order of a few nanoseconds. This delay was incorporated in the timing of colliding events by delaying transmission of the event by $\mathrm{p_e}\times\mathrm{t_{arb}}$ where $\mathrm{p_e}$ is the event priority decided by the arbiter. Thus, the estimated arbitration delay was fed to the processing pipeline along with the recorded neural signals and, calibrated thresholds for ON/OFF pulse generation and spike detection ($\mathrm{Th_{ON}}$/ $\mathrm{Th_{OFF}}$ and $\mathrm{Th_{SPD}}$ respectively). To evaluate the correctness of neural signal encoding in the form of neural events using the proposed scheme, we recover the signal by following the steps detailed in Section - \ref{sec:SignalRecovery}. The recovered signal which approximates the original channel recording was thus obtained. The drift in the trace of the recovered signal from the baseline owing to the open loop asynchronous behavior of delta modulation was removed using a high-pass filter on the recovered signal. The following subsections elaborate on the neural recording datasets that were used for the simulations in this work, along with the evaluation metrics used and key results.
\subsection{Dataset}
The simulations in this work have been performed for three diverse datasets, ranging from single-channel synthetic data to 100-channel non-human primate (NHP) motor cortex and 384-channel mice visual cortex neural recordings. The details of the datasets used are as follows:
\begin{itemize}
    \item Synthetic dataset provided in \cite{syntheticDataset} with varying noise levels ($\mathrm{\sigma_{noise}}$ = 0.05, 0.1, 0.15 and 0.2) and sampling frequency of $24$ KHz with spike duration of $\approx 1-2$ ms. The dataset also contains ground truth for the spike detection task. For worst-case simulation of large n-channel electrode arrays, the single-channel signal is replicated to all the n-channels.
    \item Non-human primate (NHP) motor cortex recordings containing 100 channels sampled at 30 KHz previously used in \cite{ShoebDatou,camilo_plos} for decoding motor intention. Since no ground truth is available for this dataset, spike detection was performed on the original recorded signal using the absolute threshold method in \cite{syntheticDataset}. The resultant spike detections were used as a ground truth of potential spike samples for performance analysis of the different compression methods investigated in this work.
    \item Recordings of mice's visual cortex to visual stimuli with 384 channels Neuropixel \cite{Lopez2016} at $30$~KHz and $<7~\mu$V RMS noise levels were made available in \cite{miceCortexDataset}. Spike detection ground truth containing potential spike times was obtained in the same way as was described for the NHP dataset above. Spike signal duration is $<1~\mu$ S in the Neuropixel recording.
\end{itemize}

\subsection{Evaluation Metrics} \label{sec:evaluation_metrics}
%We evaluate the tradeoff between compression and task accuracy using spike shape reconstruction error as a metric. We also show that better tradeoffs may be obtained by using other metrics, such as spike detection accuracy, which are kept for further exploration in the future.
In order to analyze the effect of neuromorphic compression on iBMI we evaluate the spike information retention i.e., how well the compressed signal preserves spike shape (used in tasks such as spike sorting). This is measured in terms of signal recovery performance. We also show that better tradeoffs may be obtained by using other metrics, such as task-specific performance analysis. To this end, we use spike detection performance to assess the task-specific usability of signal recovered from the compressed data for downstream iBMI tasks, such as spike-based decoding. 
%The calculations in the following sections assume a $100\times100$ neural recording array.
\subsubsection{Signal Recovery} \label{sec:SignalRecovery}
Signal recovery was performed by stair-step reconstruction of the transmitted event pulses. We recover the signal from APM event packets by adding/subtracting the $\mathrm{Th_{ON}}$/$\mathrm{Th_{OFF}}$ value at the corresponding ON ($\mathrm{+1}$) and OFF ($\mathrm{+1}$) event times. Signal recovery was done for PCM packets by adding/subtracting $\mathrm{Th_{ON}}$/$\mathrm{Th_{OFF}}$ multiplied by the event counts ($\mathrm{n_{ON}}$/$\mathrm{n_{OFF}}$) in the event packets. The recovered signal was then resampled to match the sampling rate of the neural recording, as shown in Fig. \ref{fig:Figure4}(a). To evaluate the signal recovery performance, we used two common metrics - root-mean-square error (RMSE) and Pearson's correlation coefficient (CC). RMSE was normalized to $[0,1]$. These metrics are indicative of spike shape preservation in the event data obtained from the proposed APM/PCM pipeline. A Low RMSE and high CC indicate good spike shape preservation or high recovery, i.e., the recovered signal closely approximates the original signal and may be assumed to yield good performance in tasks such as spike cell classification/clustering that depend on spike shape. These metrics were computed between the original signal and the signal recovered from the event pulse packets. 

\subsubsection{Spike Detection (SPD)}\label{sec:spd}
Evaluation of spike detection performance is done by determining detection accuracy (A), sensitivity (S), and false detection rate (FDR) as shown in Eq. \ref{eq:metrics} \cite{zzSPD}.
\begin{gather}
\label{eq:metrics}
\mathrm{S} =\frac{\mathrm{TP}}{\mathrm{TP + FN}};~ \mathrm{FDR} =\frac{\mathrm{FP}}{\mathrm{TP + FP}};~ 
\mathrm{A} =\frac{\mathrm{TP}}{\mathrm{TP + FP + FN}}
\end{gather}
Spike detections occurring within a tolerance window of $\mathrm{t_{spike}\pm\delta_{\mathrm{tolerance}}}$ ($\delta_{\mathrm{tolerance}}$ here is about half the spike duration, i.e., $\approx0.5$~ms) are marked as true positives ($\mathrm{TP}$), spurious detections that are absent in the ground truth are marked as false positives ($\mathrm{FP}$) and the missed detections are marked as false negatives ($\mathrm{FN}$). Accuracy ($\mathrm{A}$), sensitivity ($\mathrm{S}$), and false detection rate ($\mathrm{FDR}$) calculated using Eq. (\ref{eq:metrics}) are used as metrics for measuring spike detection performance.
Accuracy is an overall metric because it takes into account sensitivity and FDR. A high sensitivity is desired for SPD performance evaluation in firing rate based BMI systems, especially considering the sparsity of spikes. Since the synthetic dataset used contains ground truth for SPD, we performed SPD using both - the absolute threshold crossing method (AT-SPD) \cite{syntheticDataset} and non-linear energy operator (NEO-SPD)\cite{NEO} method involving threshold crossing based detection on the NEO-enhanced signal. In the calibration stage, the spike detection threshold for NEO-SPD was determined as $\mathrm{Th_{SPD}} = 8 \times median($NEO(neuralSignal)$)$.
Spikes are detected for all NEO(recoveredSignal) $>$ $\mathrm{Th_{SPD}}$. For the signal recovered from events of the synthetic dataset, NEO-SPD and AT-SPD yielded similar SPD performance results. Therefore, for simplicity, AT-SPD was used for `potential spike times' ground truth generation from NHP and Neuropixel datasets, and for the evaluation of SPD performance evaluation on the signal recovered from the events stream corresponding to these datasets.

\subsection{Choice of threshold}\label{sec:threshold}
\begin{figure}[t]
\centering    
\includegraphics[width=0.98\columnwidth]{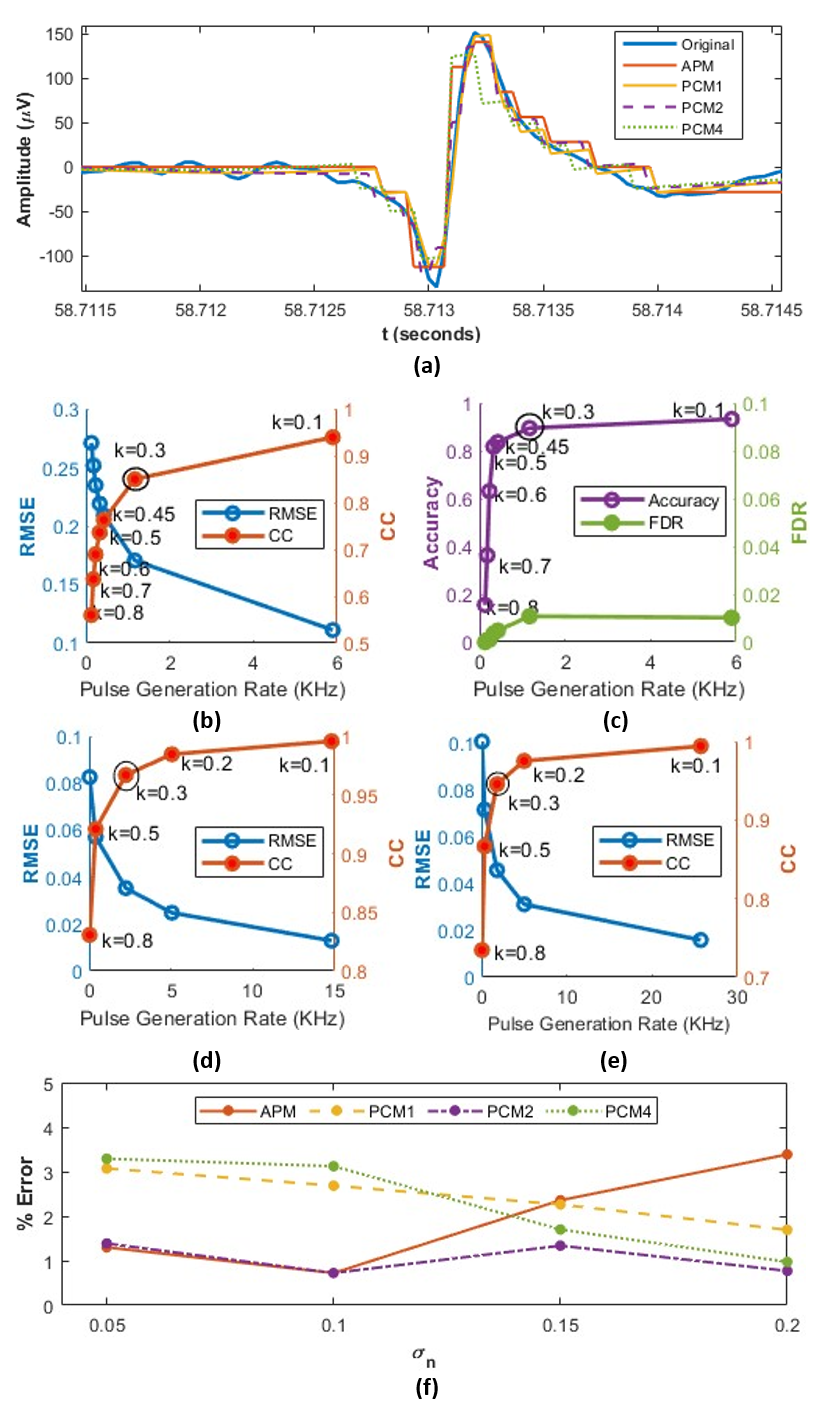}
 \caption{(a) Snippet of signal recovery for APM and PCM-1, 2, and 4 compared to the original neural signal shown by the blue line. (b) Illustration of choice of threshold from trade-off curves for PCM1 for the tasks of signal reconstruction and (c) spike detection for signal channel synthetic dataset. Optimal input referred thresholds  $\pm k\times  \mathrm{V_{spike-max}}$ for $k=\pm0.3$ and $\pm0.45$ for the two cases. Similar trade-off curves are obtained for choosing the threshold for the (d) 100-channel NHP and (e) 384-channel Neuropixel datasets. (f) Percentage error between the theoretically estimated TDR and the TDR obtained from the simulations.}
\label{fig:Figure4}
\end{figure}

For APM and PCM, the ON and OFF thresholds were determined from trade-off curves between threshold and performance metrics mentioned in \ref{sec:evaluation_metrics}. During the calibration stage (a few seconds long), the input-referred pulse generation threshold for each recording channel was determined as a factor of the spike amplitudes $\mathrm{V_{spike-max}}$ and is obtained as follows:
\begin{align}
\label{eq:threshold}
    \mathrm{Th_{ON/OFF}}=\pm k \times \mathrm{V_{spike-max}}
\end{align}
The factor $k$ was obtained from trade-off curves as shown in Fig. \ref{fig:Figure4}(b) by sweeping $k$ in the range $(0,1)$ in steps of $0.1$ and determining the value at which $k$ balances off the performance metrics and pulse generation rate. As shown in the left plot of Fig. \ref{fig:Figure4}(b), for the synthetic dataset, an input referred threshold obtained with of $k=\pm0.3$ was found to provide a good tradeoff between data rate and signal recovery. As expected, spike detection required a less stringent threshold, obtained with  $k=\pm0.45$. $\mathrm{Th_{ON/OFF}}$ is a parameter of the asynchronous delta modulator and therefore affects the pulse generation rate. A larger  $\mathrm{Th_{ON/OFF}}$ would result in a lower data rate. However, this results in coarser reconstruction of the signal from the pulse data and suppression of samples whose values are below the threshold, resulting in degraded signal recovery (affecting CC and RMSE) and spike detection performance (affecting Acc and S). Similar trade-off curves were obtained for the NHP and Neuropixel datasets, and a $k=\pm0.3$ yielded better performance without a significant increase in the data rate as shown in Fig. \ref{fig:Figure4}(d-e). The more stringent threshold for signal recovery is used to report the results presented in the following subsections.

\subsection{Data rates for the different modes}
The theoretical model for the estimation of transmission data rate (TDR) in terms of the firing rate of the biological neuron, for the architecture studied in this work is:
\newcommand\numberthis{\addtocounter{equation}{1}\tag{\theequation}}
%\mathrm{R_p} &= \mathrm{R_{AP}} + \mathrm{R_{noise}} = \mathrm{f_{neu}} \times \mathrm{N_{AP-pulses}} +  \mathrm{R_{noise}}\notag\\
\begin{align*}
\label{eq:datarate}
\mathrm{TDR}&= (\mathrm{N_r} \times \mathrm{N_c}) \times \mathrm{R_p} \times (\mathrm{n_b} + \log_{2}\mathrm{N_r} + \log_{2}\mathrm{N_c} ) \notag\\
\mathrm{R_p} &= \begin{cases} 
\mathrm{f_{neu}} \times \mathrm{N_{AP}} +  \mathrm{R_{noise}}, &\text{for~APM}\\
\mathrm{R_{bin}}, &\text{for~PCM} \\
\mathrm{f_{neu}}\times \mathrm{N_{spike}}\, &\text{for~SPDWOR}\\
\mathrm{f_s}, &\text{for~\cite{19584Electrodes,5120Electtrodes,59760Electrodes}}\\
\end{cases}
\\
\mathrm{n_b} &= \begin{cases} 
1, &\text{for~APM} \\
\mathrm {n_{b-ON}}+\mathrm{n_{b-OFF}}, &\text{for~PCM} \\
\mathrm{b_{ADC}}, &\text{for~SPDWOR,\cite{19584Electrodes,5120Electtrodes,59760Electrodes}} \numberthis
\end{cases}
\end{align*}
where $\mathrm{R_p}$ is the sample/pulse generation rate for an $\mathrm{N_r}\times \mathrm{N_c}$ array requiring $\mathrm{n_b}$ bits to represent the pulse, $\mathrm{f_{neu}}$ is the biological spike firing rate, $\mathrm{N_{AP}}$ is the average number of pulses generated per spike, $\mathrm{n_b}$ is the number of bits per pulse/sample, $\mathrm{R_{noise}}$ is the pulse generation rate corresponding to non-spike samples. $\mathrm{R_{bin}}$ is the rate of non-empty event count bins in PCM, which depends on the number of bins per second ($\mathrm{n_b}$) and the probability of non-empty bin ($\alpha_\mathrm{b}$), and is related as follows:
\begin{align}
\label{eq:rbin}
\mathrm{R_{bin}} = \alpha_\mathrm{b}\times\mathrm{n_b}
\end{align}
$\alpha_\mathrm{b}$ in Eq. \ref{eq:rbin} represents the sparsity of PCM event counts. SPDWOR involves transmitting $\mathrm{b_{ADC}}$ bits for $\mathrm{f_{neu}}\times \mathrm{N_{spike}}$ spike samples, where $\mathrm{N_{spike}}$ is the number of samples per spike. The theoretical model was validated by comparing with several seconds of synthetic data and was found to match well with an error of $\pm 5\%$.
\begin{figure}[t]
\centering    
\includegraphics[width=0.48\textwidth]{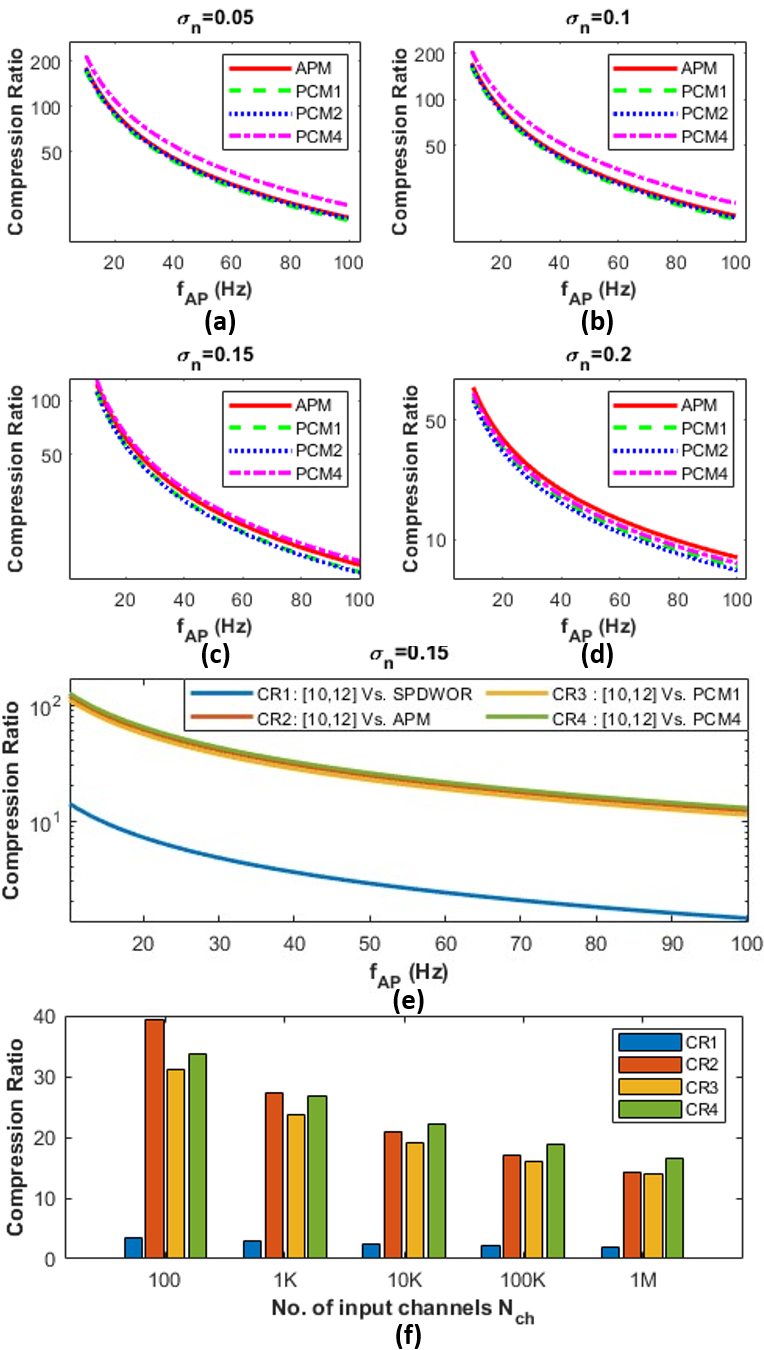}
 \caption{(a)-(d) Comparison of CR for APM and PCM on datasets with varying noise levels and neural firing rates. (e) Analysis of CR with increasing neural firing rates and comparison with different compression methods (f) Analysis of CR with increasing electrode count.}
\label{fig:compressionPlots}
\end{figure}

\subsection{Compression ratio}
The extent of compression for the proposed method is evaluated by computing the compression ratio (CR), which is defined as the ratio of the transmission data rate for full sample transmission ($\mathrm{TDR_{fs}}$) as done in \cite{19584Electrodes,5120Electtrodes,59760Electrodes} to the transmission data rate of spike-sample transmission ($\mathrm{TDR_{spk}}$) as done in SPDWOR \cite{SPDWOR} or that of the proposed PCM/APM ($\mathrm{TDR_{APM/PCMn}}$). The CRs for different transmission modes (APM and PCM) were computed using the following equations:
\begin{align}
\label{eq:cr}
\mathrm{CR1} &=\frac{\mathrm{TDR_{fs}}}{\mathrm{TDR_{spk}}}&;~ \mathrm{CR2} &=\frac{\mathrm{TDR_{fs}}}{\mathrm{TDR_{APM}}};\\\notag
\mathrm{CR3} &=\frac{\mathrm{TDR_{fs}}}{\mathrm{TDR_{PCM1}}}&;~
\mathrm{CR4} &=\frac{\mathrm{TDR_{fs}}}{\mathrm{TDR_{PCM4}}}
\end{align}
The CR comparison plot shown in Fig~\ref{fig:compressionPlots}~(a-d) was obtained for the synthetic dataset by sweeping the firing rate $\mathrm{f_{neu}}$ of the channel and linearly extrapolating the compression ratio for different $\mathrm{f_{neu}}$ for different background noise levels. As expected, the $\mathrm{TDR_{APM/PCMn}}$ increases with increasing noise levels (due to added events generated by background noise) and results in a drop in the CR. Higher the $\mathrm{f_{neu}}$ higher is the APM/PCM event generation rate ($\mathrm{R_p}$ in Eq. \ref{eq:datarate}), resulting in a drop in the CR as $\mathrm{f_{neu}}$ increases. An ideal SPDWOR-like implementation transmits all the samples (8-12 bits per sample besides the address bits) corresponding to each of the spikes, irrespective of the change in the signal value compared to the previously recorded sample. In contrast, APM transmits 1 bit and PCM transmits  $\mathrm{n_{ON}+n_{OFF}}$ (typically $ < 12$) bits only when the signal exceeds $\mathrm{Th_{ON/OFF}}$, thereby transmitting fewer bits per spike compared to SPDWOR. Therefore, at neural firing rates of $\approx$ $30-60$~Hz, the CR of the proposed method is $\approx20-50$ and $6-8\times$ SPDWOR \cite{SPDWOR} as shown in Fig.\ref{fig:compressionPlots}(e). Even though CR in proposed methods decreases with increasing electrode numbers (Fig. \ref{fig:compressionPlots}(f)) due to a higher number of bits needed to encode the address, it is still $\approx 3\times$ better than \cite{SPDWOR} for $10$K channels. For the 100-channel NHP dataset with an average neural firing rate per channel of $\approx 62$~Hz, CR1 = $3.23$, CR2 = $25.2$, CR3 = $15.4$, and CR4 = $850$ were obtained, which is consistent with the estimated CR in Fig. \ref{fig:compressionPlots}(f).
% \begin{align}
% \label{eq:compressionRatio}
% CR1 &= \frac{DR_{All samples}}{DR_{All generated pulses}} \notag \\
% CR2 &= DR_{Spike samples} / DR_{All generated pulses}\notag \\
% CR2 &= DR_{Spike samples} / DR_{Spike pulses}
% \end{align}
% Please add the following required packages to your document preamble:
% \usepackage{multirow}
\begin{table*}[t]
\caption{Analysis of the effect of neuromorphic compression. The best results for each of the criteria are highlighted in boldface.}
\centering
\resizebox{0.95\textwidth}{!}{
% Please add the following required packages to your document preamble:
% \usepackage{multirow}
\begin{tabular}{|ccccccccc|}
\hline
\multicolumn{9}{|c|}{Threshold     = $\pm0.3\times \mathrm{V_{spike-max}}$}                                                                                                                                                                                                                                                                                                     \\ \hline
\multicolumn{1}{|c|}{Dataset}                                              & \multicolumn{1}{c|}{\#Channels}  & \multicolumn{1}{c|}{Mode} & \multicolumn{1}{c|}{RMSE}   & \multicolumn{1}{c|}{CC}     & \multicolumn{1}{c|}{A}   & \multicolumn{1}{c|}{S}     & \multicolumn{1}{c|}{FDR}  & DR (Mbps) \\ \hline
\multicolumn{1}{|c|}{\multirow{4}{*}{Synthetic ($\mathrm{\sigma_{noise}}$ = 0.05)}} & \multicolumn{1}{c|}{\multirow{4}{*}{10K}} & \multicolumn{1}{c|}{APM}           & \multicolumn{1}{c|}{\textbf{0.1054}} & \multicolumn{1}{c|}{\textbf{0.923}}  & \multicolumn{1}{c|}{92.82}          & \multicolumn{1}{c|}{92.82}          & \multicolumn{1}{c|}{\textbf{0}}    & 76.75              \\ \cline{3-9} 
\multicolumn{1}{|c|}{}                                                              & \multicolumn{1}{c|}{}                     & \multicolumn{1}{c|}{PCM1}          & \multicolumn{1}{c|}{0.1236}          & \multicolumn{1}{c|}{0.8937}          & \multicolumn{1}{c|}{\textbf{99.31}} & \multicolumn{1}{c|}{\textbf{99.86}} & \multicolumn{1}{c|}{0.55}          & 80.22              \\ \cline{3-9} 
\multicolumn{1}{|c|}{}                                                              & \multicolumn{1}{c|}{}                     & \multicolumn{1}{c|}{PCM2}          & \multicolumn{1}{c|}{0.13}            & \multicolumn{1}{c|}{0.8827}          & \multicolumn{1}{c|}{99.04}          & \multicolumn{1}{c|}{99.57}          & \multicolumn{1}{c|}{0.53}          & 78                 \\ \cline{3-9} 
\multicolumn{1}{|c|}{}                                                              & \multicolumn{1}{c|}{}                     & \multicolumn{1}{c|}{PCM4}          & \multicolumn{1}{c|}{0.7268}          & \multicolumn{1}{c|}{0.2681}          & \multicolumn{1}{c|}{71.23}          & \multicolumn{1}{c|}{71.81}          & \multicolumn{1}{c|}{1.13}          & \textbf{63.59}     \\ \hline
\multicolumn{1}{|c|}{\multirow{4}{*}{Synthetic ($\mathrm{\sigma_{noise}}$ = 0.1)}}  & \multicolumn{1}{c|}{\multirow{4}{*}{10K}} & \multicolumn{1}{c|}{APM}           & \multicolumn{1}{c|}{\textbf{0.1196}} & \multicolumn{1}{c|}{\textbf{0.9077}} & \multicolumn{1}{c|}{93.58}          & \multicolumn{1}{c|}{93.58}          & \multicolumn{1}{c|}{\textbf{0}}    & 81.12              \\ \cline{3-9} 
\multicolumn{1}{|c|}{}                                                              & \multicolumn{1}{c|}{}                     & \multicolumn{1}{c|}{PCM1}          & \multicolumn{1}{c|}{0.1386}          & \multicolumn{1}{c|}{0.8752}          & \multicolumn{1}{c|}{\textbf{97.68}} & \multicolumn{1}{c|}{\textbf{99.72}} & \multicolumn{1}{c|}{2.06}          & 85.51              \\ \cline{3-9} 
\multicolumn{1}{|c|}{}                                                              & \multicolumn{1}{c|}{}                     & \multicolumn{1}{c|}{PCM2}          & \multicolumn{1}{c|}{0.1401}          & \multicolumn{1}{c|}{0.8729}          & \multicolumn{1}{c|}{96.71}          & \multicolumn{1}{c|}{97.93}          & \multicolumn{1}{c|}{1.27}          & 83.19              \\ \cline{3-9} 
\multicolumn{1}{|c|}{}                                                              & \multicolumn{1}{c|}{}                     & \multicolumn{1}{c|}{PCM4}          & \multicolumn{1}{c|}{0.4165}          & \multicolumn{1}{c|}{0.4506}          & \multicolumn{1}{c|}{70.47}          & \multicolumn{1}{c|}{71.17}          & \multicolumn{1}{c|}{1.37}          & \textbf{67.19}     \\ \hline
\multicolumn{1}{|c|}{\multirow{4}{*}{Synthetic ($\mathrm{\sigma_{noise}}$ = 0.15)}} & \multicolumn{1}{c|}{\multirow{4}{*}{10K}} & \multicolumn{1}{c|}{APM}           & \multicolumn{1}{c|}{\textbf{0.1368}} & \multicolumn{1}{c|}{\textbf{0.8903}} & \multicolumn{1}{c|}{93.99}          & \multicolumn{1}{c|}{94.09}          & \multicolumn{1}{c|}{\textbf{0.12}} & 114.22             \\ \cline{3-9} 
\multicolumn{1}{|c|}{}                                                              & \multicolumn{1}{c|}{}                     & \multicolumn{1}{c|}{PCM1}          & \multicolumn{1}{c|}{0.1559}          & \multicolumn{1}{c|}{0.8561}          & \multicolumn{1}{c|}{\textbf{95.53}} & \multicolumn{1}{c|}{\textbf{96.45}} & \multicolumn{1}{c|}{0.99}          & 124.86             \\ \cline{3-9} 
\multicolumn{1}{|c|}{}                                                              & \multicolumn{1}{c|}{}                     & \multicolumn{1}{c|}{PCM2}          & \multicolumn{1}{c|}{0.156}           & \multicolumn{1}{c|}{0.8563}          & \multicolumn{1}{c|}{89.15}          & \multicolumn{1}{c|}{90.32}          & \multicolumn{1}{c|}{1.43}          & 126.25             \\ \cline{3-9} 
\multicolumn{1}{|c|}{}                                                              & \multicolumn{1}{c|}{}                     & \multicolumn{1}{c|}{PCM4}          & \multicolumn{1}{c|}{0.2344}          & \multicolumn{1}{c|}{0.7117}          & \multicolumn{1}{c|}{71.84}          & \multicolumn{1}{c|}{72.88}          & \multicolumn{1}{c|}{1.95}          & \textbf{108.66}    \\ \hline
\multicolumn{1}{|c|}{\multirow{4}{*}{Synthetic ($\mathrm{\sigma_{noise}}$ = 0.2)}}  & \multicolumn{1}{c|}{\multirow{4}{*}{10K}} & \multicolumn{1}{c|}{APM}           & \multicolumn{1}{c|}{\textbf{0.1459}} & \multicolumn{1}{c|}{\textbf{0.892}}  & \multicolumn{1}{c|}{\textbf{92.02}} & \multicolumn{1}{c|}{\textbf{93.23}} & \multicolumn{1}{c|}{1.4}           & 178.24             \\ \cline{3-9} 
\multicolumn{1}{|c|}{}                                                              & \multicolumn{1}{c|}{}                     & \multicolumn{1}{c|}{PCM1}          & \multicolumn{1}{c|}{0.1705}          & \multicolumn{1}{c|}{0.8499}          & \multicolumn{1}{c|}{89.5}           & \multicolumn{1}{c|}{90.38}          & \multicolumn{1}{c|}{\textbf{1.08}} & 202.29             \\ \cline{3-9} 
\multicolumn{1}{|c|}{}                                                              & \multicolumn{1}{c|}{}                     & \multicolumn{1}{c|}{PCM2}          & \multicolumn{1}{c|}{0.1696}          & \multicolumn{1}{c|}{0.8521}          & \multicolumn{1}{c|}{80.29}          & \multicolumn{1}{c|}{81.3}           & \multicolumn{1}{c|}{1.52}          & 212.26             \\ \cline{3-9} 
\multicolumn{1}{|c|}{}                                                              & \multicolumn{1}{c|}{}                     & \multicolumn{1}{c|}{PCM4}          & \multicolumn{1}{c|}{0.2468}          & \multicolumn{1}{c|}{0.7183}          & \multicolumn{1}{c|}{71.19}          & \multicolumn{1}{c|}{71.91}          & \multicolumn{1}{c|}{1.38}          & \textbf{192.54}    \\ \hline
\multicolumn{1}{|c|}{\multirow{4}{*}{NHP}}                                          & \multicolumn{1}{c|}{\multirow{4}{*}{100}} & \multicolumn{1}{c|}{APM}           & \multicolumn{1}{c|}{\textbf{0.0983}} & \multicolumn{1}{c|}{\textbf{0.8997}} & \multicolumn{1}{c|}{\textbf{90.28}} & \multicolumn{1}{c|}{\textbf{96.91}} & \multicolumn{1}{c|}{\textbf{6.7}}  & 1.19               \\ \cline{3-9} 
\multicolumn{1}{|c|}{}                                                              & \multicolumn{1}{c|}{}                     & \multicolumn{1}{c|}{PCM1}          & \multicolumn{1}{c|}{0.1318}          & \multicolumn{1}{c|}{0.8515}          & \multicolumn{1}{c|}{87.32}          & \multicolumn{1}{c|}{93.2}           & \multicolumn{1}{c|}{8}             & 1.95               \\ \cline{3-9} 
\multicolumn{1}{|c|}{}                                                              & \multicolumn{1}{c|}{}                     & \multicolumn{1}{c|}{PCM2}          & \multicolumn{1}{c|}{0.1443}          & \multicolumn{1}{c|}{0.8234}          & \multicolumn{1}{c|}{86.51}          & \multicolumn{1}{c|}{92.58}          & \multicolumn{1}{c|}{0.08}          & 0.36               \\ \cline{3-9} 
\multicolumn{1}{|c|}{}                                                              & \multicolumn{1}{c|}{}                     & \multicolumn{1}{c|}{PCM4}          & \multicolumn{1}{c|}{0.2311}          & \multicolumn{1}{c|}{0.7422}          & \multicolumn{1}{c|}{65.01}          & \multicolumn{1}{c|}{88.38}          & \multicolumn{1}{c|}{0.29}          & \textbf{0.04}      \\ \hline
\multicolumn{1}{|c|}{\multirow{4}{*}{Neuropixel}}                                    & \multicolumn{1}{c|}{\multirow{4}{*}{384}} & \multicolumn{1}{c|}{APM}           & \multicolumn{1}{c|}{\textbf{0.0881}} & \multicolumn{1}{c|}{\textbf{0.8981}} & \multicolumn{1}{c|}{\textbf{84.94}} & \multicolumn{1}{c|}{89.45}          & \multicolumn{1}{c|}{\textbf{5.54}} & 6.58               \\ \cline{3-9} 
\multicolumn{1}{|c|}{}                                                              & \multicolumn{1}{c|}{}                     & \multicolumn{1}{c|}{PCM1}          & \multicolumn{1}{c|}{0.088}           & \multicolumn{1}{c|}{0.9005}          & \multicolumn{1}{c|}{74.39}          & \multicolumn{1}{c|}{\textbf{91.04}} & \multicolumn{1}{c|}{19.63}         & 10.99              \\ \cline{3-9} 
\multicolumn{1}{|c|}{}                                                              & \multicolumn{1}{c|}{}                     & \multicolumn{1}{c|}{PCM2}          & \multicolumn{1}{c|}{0.1186}          & \multicolumn{1}{c|}{0.854}           & \multicolumn{1}{c|}{71.36}          & \multicolumn{1}{c|}{89.63}          & \multicolumn{1}{c|}{22.15}         & 2.03               \\ \cline{3-9} 
\multicolumn{1}{|c|}{}                                                              & \multicolumn{1}{c|}{}                     & \multicolumn{1}{c|}{PCM4}          & \multicolumn{1}{c|}{0.1922}          & \multicolumn{1}{c|}{0.7201}          & \multicolumn{1}{c|}{57.86}          & \multicolumn{1}{c|}{87.97}          & \multicolumn{1}{c|}{36.86}         & \textbf{1.83}      \\ \hline
\multicolumn{1}{|c|}{\multirow{4}{*}{Neuropixel}}                                   & \multicolumn{1}{c|}{\multirow{4}{*}{100}} & \multicolumn{1}{c|}{APM}           & \multicolumn{1}{c|}{0.0808}          & \multicolumn{1}{c|}{0.8389}          & \multicolumn{1}{c|}{\textbf{88.44}} & \multicolumn{1}{c|}{\textbf{94.36}} & \multicolumn{1}{c|}{\textbf{6.53}} & \textbf{0.25}      \\ \cline{3-9} 
\multicolumn{1}{|c|}{}                                                              & \multicolumn{1}{c|}{}                     & \multicolumn{1}{c|}{PCM1}          & \multicolumn{1}{c|}{\textbf{0.0678}} & \multicolumn{1}{c|}{\textbf{0.9101}} & \multicolumn{1}{c|}{72.58}          & \multicolumn{1}{c|}{91.2}           & \multicolumn{1}{c|}{21.88}         & 0.46               \\ \cline{3-9} 
\multicolumn{1}{|c|}{}                                                              & \multicolumn{1}{c|}{}                     & \multicolumn{1}{c|}{PCM2}          & \multicolumn{1}{c|}{0.0979}          & \multicolumn{1}{c|}{0.8571}          & \multicolumn{1}{c|}{72.28}          & \multicolumn{1}{c|}{89.62}          & \multicolumn{1}{c|}{21.05}         & 0.48               \\ \cline{3-9} 
\multicolumn{1}{|c|}{}                                                              & \multicolumn{1}{c|}{}                     & \multicolumn{1}{c|}{PCM4}          & \multicolumn{1}{c|}{0.2119}          & \multicolumn{1}{c|}{0.6238}          & \multicolumn{1}{c|}{63.39}          & \multicolumn{1}{c|}{86.17}          & \multicolumn{1}{c|}{28.93}         & 0.42               \\ \hline
\end{tabular}

} \label{Table:CompresssionEffect}
\end{table*}

\subsection{Effect of compression}
As discussed in Section \ref{sec:evaluation_metrics}, signal recovery and spike detection metrics are used to understand the effect of compression. Table \ref{Table:CompresssionEffect} summarizes the effect of neuromorphic compression on large iBMI for the two modes of pulse transmission- APM and PCM for bin widths that are 1,2 and 4 times (PCM1, PCM2, and PCM3) the sampling interval used in conventional neural signal transmission for all the three datasets introduced earlier. Over 90\% of the spike shape can be fully recovered on average, from APM, whereas the recovered signal is relatively degraded with increasing levels of compression in PCM, as seen in the RMSE and CC columns of Table \ref{Table:CompresssionEffect}. While PCM4 ensures lower TDR compared to APM, it comes at the cost of spike shape recovery. The higher the level of PCM, the coarser the event generation and recovery of the signal.

Spike detection is a typical but significant task in firing-rate based iBMI systems. Therefore, it is necessary to assess whether the proposed APM/PCM effectively captures the spike-time information. We do this by performing spike detection on the signal recovered from the APM/PCM events, comparing the detection from the recovered signal with the ground truth, and quantitatively computing the metrics (A, S and FDR) described in Section \ref{sec:spd}, it can also be observed that the sensitivity of spike detection is not significantly affected by the poor spike shape recovery. A higher sensitivity ensures that spikes are not missed, especially those closer to the noise level, and this factor is all the more significant in noisy channels. It can be seen that spike detection on signals recovered from APM, PCM1, and PCM2 results in very high sensitivity. As expected, spike detection performance, particularly FDR and accuracy, degrades with increasing noise ($0.05$ to $2.0$) and compression (APM to PCM4). 

For the NHP and Neuropixel datasets, the evaluation of performance metrics relied on how `potential spikes' were defined in the absence of the ground truth. We determined the absolute threshold for spike detection, which is a few factors higher than the noise margin as typically done in AT-SPD, and recorded the time of the positive and negative spike-detection threshold ($\mathrm{Th_{SPD}}$) crossing as potential spike times. Despite the constraint, the sensitivity of SPD and CC are decent for the NHP and Neuropixel datasets. We found that the spike information (spike time and shape) is well-preserved in the proposed APM/PCM. We leave spike detection directly using the generated APM/PCM events as future work.

\subsection{Collision Analysis}
\begin{figure}[t]
\centering 
\includegraphics[width=0.48\textwidth]{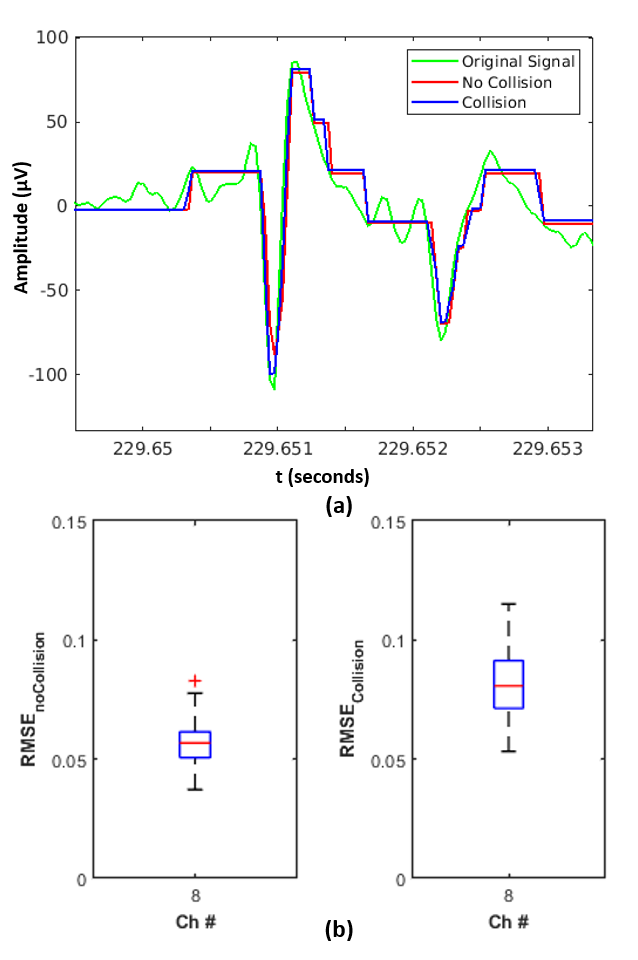}
 \caption{(a) Comparison of collision and no-collision scenarios (b) comparison of error due to collision}
\label{fig:collisionPlots}
\end{figure}

Simulations performed in this work using the synthetic datasets inherently test the robustness of the neuromorphic compression pipeline, owing to the replication of single-channel signal to $\mathrm{10K}$ channels. This results in creating a worst-case scenario where events are simultaneously generated by each of the $\mathrm{10K}$ channels, requiring arbitration without loss of events. Thanks to the sparsity of spikes and the sampling interval, the arbiter has sufficient time to complete the AER handshaking and reading out of events in time without dropping any event, which is evident from the high CC and S recorded in Table~\ref{Table:CompresssionEffect}. The AER-based read-out process however introduces a small delay owing to the arbitration time in case of collision as shown in Fig.~\ref{fig:collisionPlots}(a). For the NHP dataset, it was determined that an average of $10.0748\pm 3.5741$ channels collide in the recording, with $0 - 74$ collisions at any instance. Similarly, for the Neuropixel dataset, $29.5113 \pm 33.51$ channels collide, with $\approx5$ of them carrying spike samples. There are $0 - 321$ collisions at any instance in the Neuropixel recording.  For the NHP and Neuropixel datasets, $\approx25\%$ of events correspond to spike samples undergoing collision. If spike-sample only transmission scheme akin to SPDWOR-like read-out were to be used in scenarios like this, it would result in the loss of the spike samples that undergo collision. While SPDWOR uses collisions to an advantage to suppress samples corresponding to the background activity, it also loses some samples corresponding to the spike. The reference \cite{SPDWOR} reports $80-90\%$ spike recovery measure in terms of spike sorting performance, which indicates over a tenth of spike information is lost. The spike sample loss using the SPDWOR scheme might be more prominent in high-density neural probes that exhibit a high spatiotemporal correlation among the neighboring recording sites/channels \cite{spatioTemporalCorrealation}. Owing to the inherent noise sample suppression of the ADM and collision handling mechanism of the AER-arbiter, the loss of spike samples is negligible for the proposed APM/PCM pipeline. This can be inferred from the change in RMSE due to collision is negligible and mainly due to delay in the sample as shown in Fig.~\ref{fig:collisionPlots}(b).
\section{Discussion}
\label{sec:Discussion}
\subsection{Scalability simulation and worst-case analysis}
In the simulations for the APM/PCM pipeline using the synthetic dataset, it is to be noted that the CRs reported are for the worst-case scenario where events are simultaneously generated in each of the $\mathrm{10K}$ channels and the AER-arbiter handles the collisions. The worst-case analysis is a by-product of simulations with single-channel synthetic data of varying noise levels copied to $\mathrm{10K}$ channels. We noticed there is only a small difference in the signal recovery performance due to the arbitration delay in the proposed pipeline. Since we did not implement the SPDWOR, we estimate the data rate based on the assumption that under ideal conditions, SPDWOR captures all the samples corresponding to the spike. Work in \cite{SPDWOR} demonstrates the scalability of the wired-OR readout for up to 512-channel electrode array with about $80\%$ spike recovery, however, the increase in complexity of wiring and the effect of collision on the quality of spike information captured for higher electrode counts is still unknown.

In order to analyze the scalability of the proposed neuromorphic compression scheme for the real recordings, we take a subset of $100$ channels with the highest firing-rate channels from the 384-channel Neuropixel dataset. As seen in the last few rows of Table \ref{Table:CompresssionEffect}, TDR for APM/PCM drops when the number of channels is limited to 100 high firing-rate channels. We notice that TDR increases significantly for APM and PCM1 when scaled from 100 to 384 channels of the Neuropixel dataset while the increase is not as high for PCM2 and PCM4. The probability of non-empty bins (studied further in Section \ref{sec:sparsityOfevents}) is higher for PCM2 and PCM4, as a result, the TDR does not change dramatically with scaling. On examining the cause for higher TDR in APM/PCM1 with scaling, we identified that a significant number of events generated by an ADM in APM/PCM with a single fixed threshold ($\mathrm{Th_{ON/OFF}}$ with $k=0.3$), correspond to background activity, resulting in higher TDR (as shown in green in Fig. \ref{fig:2Th_comparisonPlot}(a)). In order to reduce TDR and improve CR, it is necessary to choose ($\mathrm{Th_{ON/OFF}}$) such that the ADM generates events only for the spike and suppresses background activity in the channel. In Section \ref{sec:further_compression} we explore some schemes to reduce TDR and improve CR by suppressing event generation due to background activity.

\begin{table}[t]
\centering
\caption{Sparsity $\alpha_b$ of PCM events}
\label{tab:sparsityOfPCM}
\resizebox{0.65\columnwidth}{!}{%
\begin{tabular}{|c|c|c|c|}
\hline
\boldmath{$\sigma_{\mathrm{noise}}$} & \textbf{PCM1} & \textbf{PCM2} & \textbf{PCM4} \\ \hline
0.05                               & 0.0256        & 0.0407        & 0.0615        \\ \hline
0.1                                & 0.03625       & 0.06165       & 0.1013        \\ \hline
0.15                               & 0.0649        & 0.11855       & 0.21045       \\ \hline
0.2                                & 0.09505       & 0.17765       & 0.31255       \\ \hline
\end{tabular}%
}
\end{table}

\subsection{Sparsity of events}\label{sec:sparsityOfevents}
While transmitting in PCM, as opposed to APM seems to have a better CR for lower noise and thresholds as seen from the results in the previous sections, the improvement in compression does not change dramatically at higher noise levels. To understand this effect, Table \ref{tab:sparsityOfPCM} presents the sparsity of PCM events, in other words, the probability of non-zero event counts in PCM bins ($\alpha_b$ in Eq. \ref{eq:rbin}). It can be seen that there is a higher probability of non-empty bins in PCM at higher noise and for lower pulse generation thresholds, which results in higher TDR and therefore lower CR. Owing to the design of the asynchronous delta modulator ADC (ADM-based pulse generator) used in this work, a fast-rising or falling input signal such as spikes results in dense pulse generation which translates to higher APM event rate and higher PCM event counts per bin, whereas a smooth/flat input signal results in few or no event pulses. Thus, the sparsity of APM events or non-zero event counts depends on the choice of the ON/OFF pulse generation threshold.

\begin{table*}[t]
\caption{Comparison of characteristics and implementation of different iBMI neural recording systems}
\label{tab:neuralRecorders_comparison}
\resizebox{\textwidth}{!}{%
\begin{tabular}{|c|c|c|c|c|c|}
\hline
\textbf{\begin{tabular}[c]{@{}c@{}}Compression Technique/\\Factor\end{tabular}} &
  \textbf{\begin{tabular}[c]{@{}c@{}}Full-sample recording\\\cite{19584Electrodes,5120Electtrodes,59760Electrodes}\end{tabular}} &
  \textbf{\begin{tabular}[c]{@{}c@{}}Spike-sample recording\\\cite{spatioTemporalCorrealation}\end{tabular}} &
  \textbf{\begin{tabular}[c]{@{}c@{}}On-chip spike sorting/\\decoding\\\cite{Liu16,Liu17,zz22,onlineSpikeSort2023,Pagin17,ShoebDatou}\end{tabular}} &
  \textbf{\begin{tabular}[c]{@{}c@{}}SPDWOR\\\cite{SPDWOR}\end{tabular}} &
  \textbf{This work} \\ \hline
Spike shape preservation &
  Y &
  Y &
  N &
  Y &
  Y \\ \hline
Circuit complexity and wiring &
  Moderate &
  High &
  High &
  High &
  Low \\ \hline
Collision handling &
  - &
  - &
  - &
  N &
  Y \\ \hline
\begin{tabular}[c]{@{}c@{}}Scalability \\ (No. of Channels)\end{tabular} &
  \begin{tabular}[c]{@{}c@{}}384 (Neuropixel)\\ 2048 (Argo)\end{tabular} &
  256 &
 64-384 &
  512 &
  \begin{tabular}[c]{@{}c@{}}10K (Synthetic)\\ 384 (Neuropixel)\end{tabular} \\ \hline
Compression ratio (CR) &
  1 &
  2-40 &
  $>$240 &
  2-20 &
  20-100 \\ \hline
\end{tabular}%
}
\end{table*}

\subsection{Further Compression}\label{sec:further_compression}
\begin{figure} [t]
\centering 
\includegraphics[width=0.98\columnwidth]{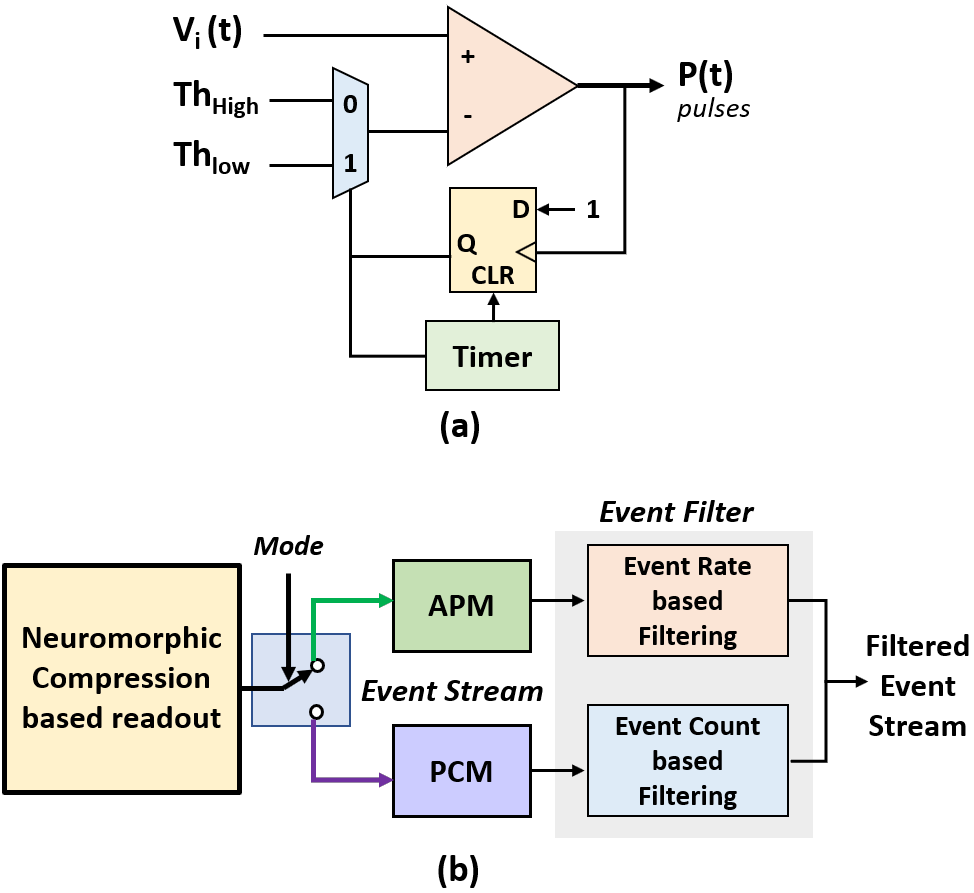}
 \caption{(a) logical block diagram of a dual threshold pulse generation asynchronous delta modulator (b) event-filters to filter out spurious events and pass events potentially corresponding to spike.}
\label{fig:2Th_comparison}
\end{figure}

\begin{table}[t]
\centering
\caption{Comparison of CR for APM using a dual threshold Asynchronous Delta Modulator at firing rate, $\mathrm{f_{AP}}=60$~Hz}
\label{tab:2th_comparison}
\resizebox{0.9\columnwidth}{!}{%
\begin{tabular}{|c|c|c|c|c|c|}
\hline
\boldmath{$\sigma_{\mathrm{noise}}$} & \boldmath{$k1$} & \boldmath{$k2$} & \textbf{TDR (Mbps)} & \textbf{CR2} & \textbf{CR2:CR1} \\ \hline
\multirow{2}{*}{0.05} & 0.6 & 0.1 & 199.9102 & 13          & 6           \\ \cline{2-6} 
                      & 0.6 & 0.3 & 57.4173  & \textbf{42} & \textbf{18} \\ \hline
\multirow{2}{*}{0.1}  & 0.6 & 0.1 & 204.856  & 12          & 5           \\ \cline{2-6} 
                      & 0.6 & 0.3 & 60.2111  & \textbf{40} & \textbf{17} \\ \hline
\multirow{2}{*}{0.15} & 0.6 & 0.1 & 214.2423 & 12          & 5           \\ \cline{2-6} 
                      & 0.6 & 0.3 & 63.0822  & \textbf{39} & \textbf{16} \\ \hline
\multirow{2}{*}{0.2}  & 0.6 & 0.1 & 225.3674 & 11          & 5           \\ \cline{2-6} 
                      & 0.6 & 0.3 & 66.2197  & \textbf{37} & \textbf{16} \\ \hline
\end{tabular}%
}

\end{table}
\begin{figure} [t]
\centering 
\includegraphics[width=0.98\columnwidth]{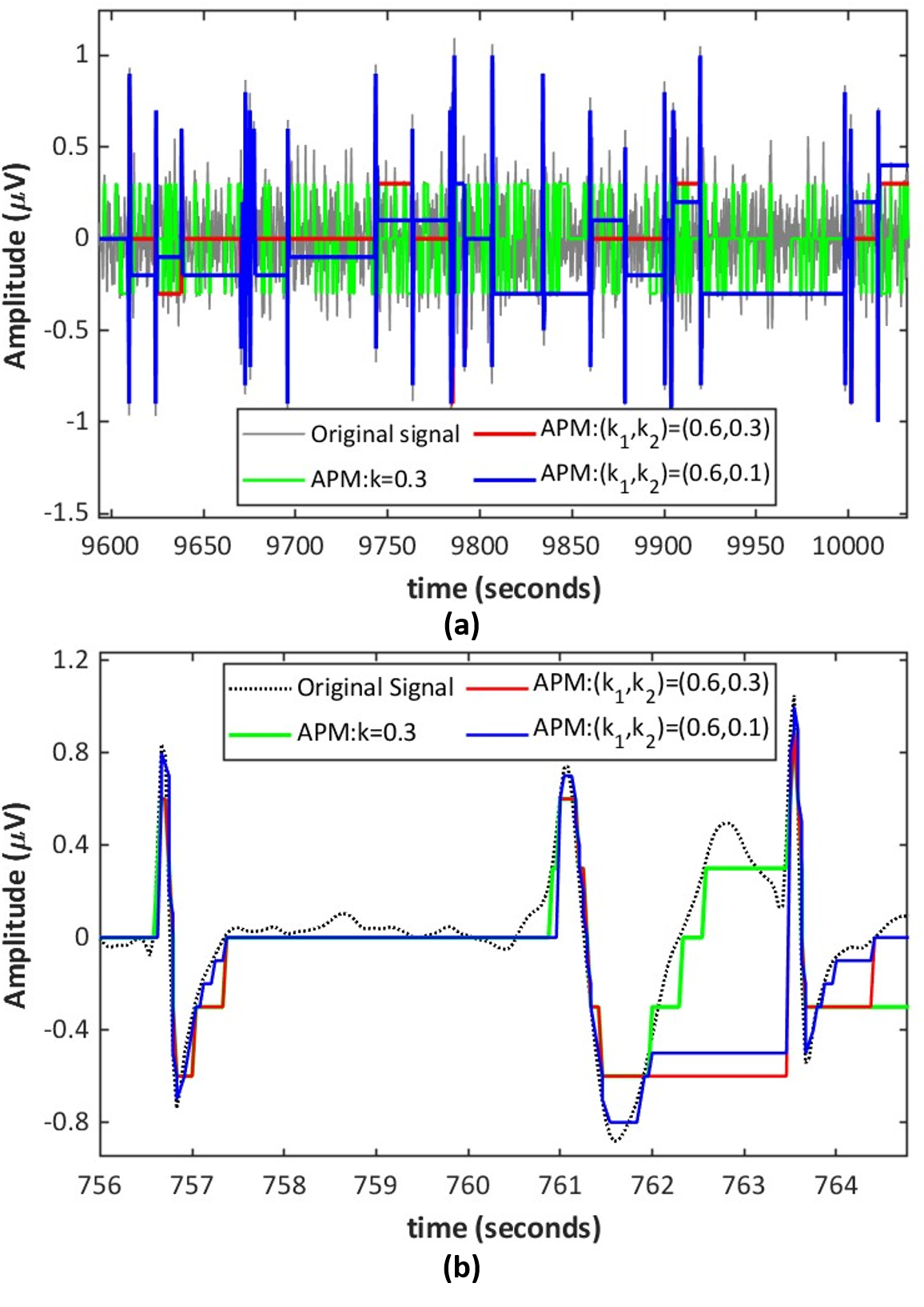}
 \caption{Comparison of single-threshold ($k=0.3$) and dual-threshold ($k1=0.6, k2=0.3/0.1$) APM recovered signal with the band-pass filtered version of the original signal. (a) A dual-threshold ADM prevents background activity from generating events (b) Depending on the choice of the factor $k_2$ for $\mathrm{Th_{Low}}$, spike shape can be captured with a coarser threshold ($k_2=0.3$) resulting in high compression or a finer threshold ($k_2=0.1$) resulting in better signal recovery performance.}
\label{fig:2Th_comparisonPlot}
\end{figure}
A good threshold for ADM is critical in ensuring higher compression. Higher thresholds i.e., increasing the factor $k$ can ensure that background activity causes no pulse/event generation and improves overall compression, however, the spike shape captured might be coarser resulting in poor recovery. An adaptive delta modulator as introduced in \cite{AdaptiveADM} or a delta modulator with a two-level threshold - one for detecting the spike and the other for finer asynchronous sampling of the spike, may be implemented. 

Fig. \ref{fig:2Th_comparison}(a) presents a logical block diagram for a simple dual threshold ADM with two sets of thresholds - $\mathrm{Th_{High}}$ and  $\mathrm{Th_{Low}}$. In this approach, the ADM initially operates with the higher threshold, until the input signal level exceeds $\mathrm{Th_{High}}$ and then switches to a lower threshold $\mathrm{Th_{Low}}$ for a fixed time $T_{timer}$ determined by the timer (typically, spike duration $\approx 1-3$~ms) to capture the spike shape with finer step size. When the timer elapses, the ADM switches back to asynchronous sampling with $\mathrm{Th_{High}}$. Following the notation used in Eq. \ref{eq:threshold}, $\mathrm{Th_{High}}$ and $\mathrm{Th_{Low}}$ are adjusted by varying the factors $k_1$ and $k_2$ respectively. Table \ref{tab:2th_comparison} summarizes the observation for two different values of $k_2$ ($k_2=0.1~\text{and}~0.3$), fixed value of $T_{timer}=1$~ms and a fixed higher $k_1=2\times$$k=0.6$ ($k=0.3$ as discussed in Section \ref{sec:threshold}). $k_2=0.3$ results in higher CR ($\approx2\times$ more than single-threshold ADM) and $16-18\times$ higher compression than SPDWOR. As shown in Fig. \ref{fig:2Th_comparisonPlot}(a), a dual-threshold ADM captures the spike shape well depending on the chosen $k_2$ and prevents spurious events from background activity that are captured by single-threshold ADM.

Another approach for increasing compression and suppressing background activity events is presented in Fig. \ref{fig:2Th_comparison}(b), where event-based temporal filters may be employed following a single-ADM neuromorphic compression based readout to limit the transmission of APM/PCM events corresponding to spikes only. Leveraging on the sparsity of events and pulse generation pattern discussed in Section \ref{sec:sparsityOfevents}, the event filters may track the event rate (for APM) or event counts (for PCM) within a temporal neighborhood to determine whether the current event corresponds to a spike or not. An event from a channel may be said to result from a spike if it is preceded by a dense spike train in the near past from the same channel. A low-complexity event-based spike detector may thus be realized as a byproduct of the event filters. However, implementation of this approach is left for future work. With an average of $4-8$ events per spike, the filtered event stream is estimated to result in up to $8\times$ more compression per channel. Thus, with future improvements to the proposed pipeline to transmit events corresponding to detected spikes only, a CR of about $50-100$ can be achieved for a mean neural firing rate of $\mathrm{f_{AP}=60}$~Hz.

\subsection{Hardware implementation}
We do not present circuit simulation of the Asynchronous Delta Modulator since such schemes have been published in several papers such as \cite{NeuromorphicRecSystem} or even the recent \cite{AdaptiveADM,NSS_PNS}. Unlike SPDWOR, no expensive wiring is needed to handle collision scenarios, which are inherently handled by the AER-based readout strategy in our proposed pipeline. Neuromorphic circuits are known for low power consumption, and thus a neuromorphic compression based iBMI system is expected to consume far less power than conventional neural recording systems with full-sample transmission. Relying on the hardware measurements from \cite{NSS_PNS} which implements an ADM-based pulse transmission in $40$~nm CMOS technology, the ADM configured consumes $\approx7~\mu$W per channel. A neuromorphic compression based readout as proposed in this work, operating in low-power mode can be estimated to result in a surface power density of $\approx4.36$~mW/cm$^2$, which lies well within the power dissipation budget of $\approx80$~mW/cm$^2$ for iBMI\cite{bmiTemperature}.

\subsection{Comparison with other works}
A summary of the high-level comparison of the proposed neuromorphic compression based neural sensing with prior works is presented in Table \ref{tab:neuralRecorders_comparison}. The proposed pipeline is a low-complexity, scalable, and high-compression neural recording system that preserves spike shape by handling collision scenarios effectively without the need for additional wiring. High compression may be obtained by recording systems that capture spike times only or perform on-chip decoding, however, this limits their adaptability to a limited set of iBMI tasks.
\section{Conclusion}
Neural electrode scaling for Nx-iBMI is severely limited by data rate and power budget. In this work, we quantitatively evaluate the effectiveness and extent of a neuromorphic compression based neural sensing architecture for iBMI, inspired by an asynchronous event detection image sensor such as the dynamic vision sensor. Transmission of pulses in APM or PCM results in compression ratios that are $2-5$ times that of transmitting spike samples as in \cite{SPDWOR} and $15-20$ times that of full sample transmission as in conventional CMOS-image sensor like readout architecture. We show that even with such high compression using the proposed neuromorphic architecture, there is $\approx 90\%$ similarity of the recovered spike shape while ensuring a spike detection accuracy of more than $92\%$ and a negligible false detection rate. A very high spike detection accuracy was obtained even for higher compression ratios of $5-50$, thus demonstrating that neuromorphic compression based neural sensing systems can perform iBMI tasks well while lowering the data rate and the transmission power of Nx-iBMI. Transmission of event pulses corresponding to spikes alone can boost the compression ratio further to $50-100$ times that of full sample transmission and $5-18$ times more compression than spike sample transmission. Future work will analyze the effects of neuromorphic compression on spike sorting and motor decoding tasks in large Nx-iBMI, and explore event-based processing for spike detection and iBMI decoding using spiking neural networks.
\section*{acknowledgment}
The work described in this paper was partially supported by a grant from the Singapore Ministry of Education Academic Research Fund Tier 2 grant (MOE-T2EP20220-0002) and the Research Grants Council of the Hong Kong Special Administrative Region, China (Project No. CityU 11200922).

%%%%%%%%%%%%%%%%%%%%%%%%%%%% REFERENCES %%%%%%%%%%%%%%%%%%%%%%%
\bibliographystyle{IEEEtran}
\bibliography{references}

%%%%%%%%%%%%%%%%%%%%%%%%%%%% BIOGRAPHY %%%%%%%%%%%%%%%%%%%%%%%
\begin{IEEEbiography}
[{\includegraphics[width=1in,height=1.25in,clip,keepaspectratio]{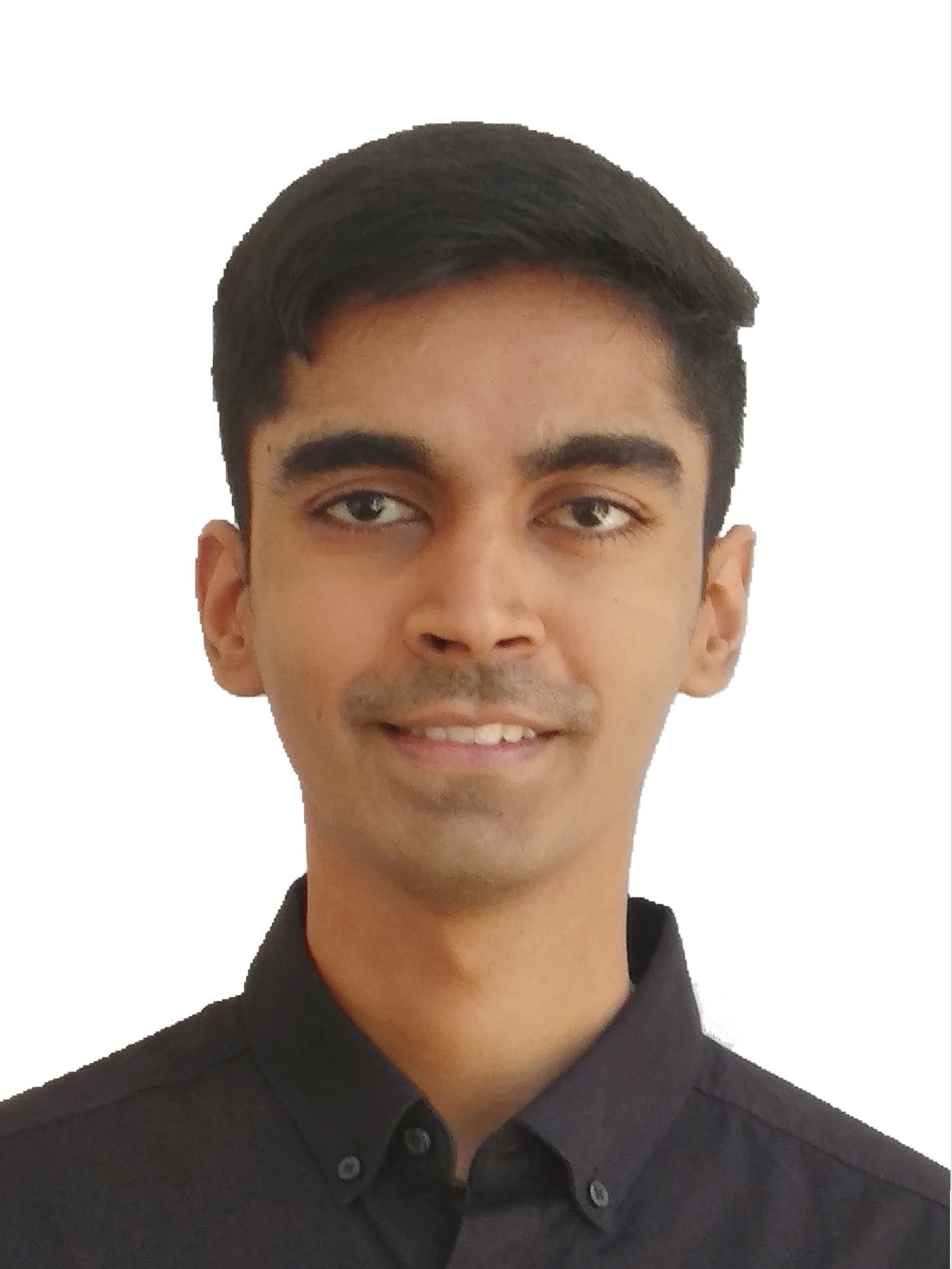}}]
{Vivek Mohan}
(S’ 14) received his B.Tech. degree in Electronics and Communication Engineering from JNTU - Hyderabad in 2016 and a Joint M.Sc. in Integrated Circuit Design from the Technical University of München and Nanyang Technological University (NTU) Singapore in 2019. After graduation, he worked as a Research Associate in the BRAIN Systems Lab - CICS, NTU in the area of neuromorphic computer vision. Vivek is currently a Ph.D. candidate at the School of Electrical and Electronic Engineering, NTU. His research interests include integrated circuit design, neurotechnology, neuromorphic engineering, and brain-machine interface.
\end{IEEEbiography}
\vspace{-1 cm}
\begin{IEEEbiography}
[{\includegraphics[width=1in,height=1.25in,clip,keepaspectratio]{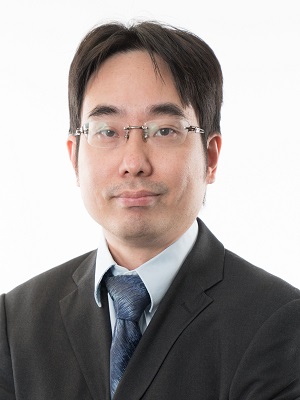}}]
{Wee Peng Tay}
(SM) received the B.S. degree in Electrical Engineering and Mathematics, and the M.S. degree in Electrical Engineering from Stanford University, Stanford, CA, USA, in 2002. He received the Ph.D. degree in Electrical Engineering and Computer Science from the Massachusetts Institute of Technology, Cambridge, MA, USA, in 2008. He is a Professor of Signal and Information Processing in the School of Electrical and Electronic Engineering at Nanyang Technological University, Singapore. Dr. Tay received the Tan Chin Tuan Exchange Fellowship in 2015. He is a co-author of the best student paper award at the Asilomar Conference on Signals, Systems, and Computers in 2012, the IEEE Signal Processing Society Young Author Best Paper Award in 2016, and the best paper award at the International Conference on Smart Power \& Internet Energy Systems in 2022. He was an Associate Editor for the IEEE Transactions on Signal Processing (2015 -- 2019) and an Editor for the IEEE Transactions on Wireless Communications (2017 -- 2023). He is currently an Associate Editor for the IEEE Transactions on Signal and Information Processing over Networks, an Associate Editor for the IEEE Internet of Things Journal, and an Editor for the IEEE Open Journal of Vehicular Technology. His research interests include signal and information processing over networks, distributed inference and estimation, statistical privacy, and robust machine learning. 
\end{IEEEbiography}
\vspace{-1 cm}
\begin{IEEEbiography}
[{\includegraphics[width=1in,height=1.25in,clip,keepaspectratio]{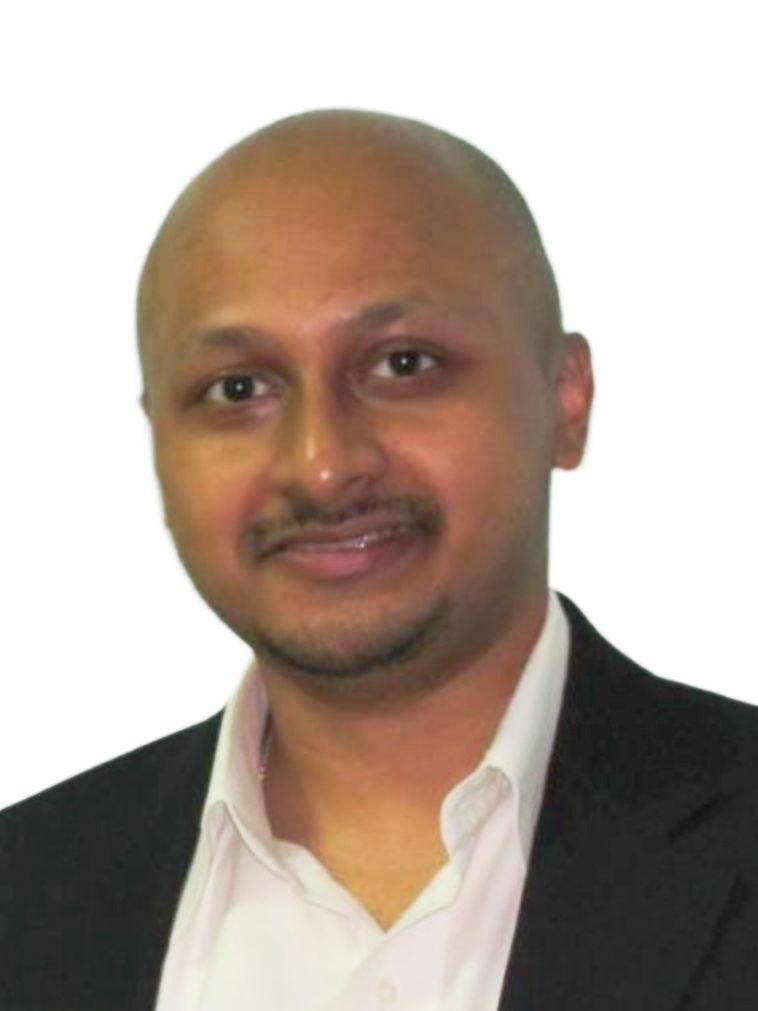}}]
{Arindam Basu}
Arindam Basu received the B.Tech and M.Tech degrees in Electronics and Electrical Communication Engineering from the Indian Institute of Technology, Kharagpur in 2005, the M.S. degree in Mathematics and PhD. degree in Electrical Engineering from the Georgia Institute of Technology, Atlanta in 2009 and 2010 respectively. Dr. Basu received the Prime Minister of India Gold Medal in 2005 from I.I.T Kharagpur. He is currently a Professor in City University of Hong Kong in the Department of Electrical Engineering and was a tenured Associate Professor at Nanyang Technological University before this. 

He is currently an Associate Editor of the IEEE Sensors journal, Frontiers in Neuroscience, IOP Neuromorphic Computing and Engineering, and IEEE Transactions on Biomedical Circuits and Systems. He has served as IEEE CAS Distinguished Lecturer for the 2016-17 period. Dr. Basu received the best student paper award at the Ultrasonics symposium, in 2006, the best live demonstration at ISCAS 2010, and a finalist position in the best student paper contest at ISCAS 2008. He was awarded MIT Technology Review's TR35 Asia Pacific award in 2012 and inducted into Georgia Tech Alumni Association's 40 under 40 class of 2022. 
\end{IEEEbiography}

% % if you will not have a photo at all:
% \begin{IEEEbiographynophoto}{John Doe}
% Biography text here.
% \end{IEEEbiographynophoto}

% % insert where needed to balance the two columns on the last page with
% % biographies
% %\newpage

% \begin{IEEEbiographynophoto}{Jane Doe}
% Biography text here.
% \end{IEEEbiographynophoto}

\end{document}